\documentclass[11pt]{article}
\newcommand{\be}{\begin{equation}}
\newcommand{\ee}{\end{equation}}
\usepackage{amsmath}
\usepackage{amssymb}
\usepackage{latexsym}
\usepackage{epsfig}

\newcommand{\ket}[1]{|#1\rangle}
\newcommand{\del}{\partial}

\newcommand{\sectiono}[1]{\section{#1}\setcounter{equation}{0}}

\begin{document}

\begin{titlepage}
% \rightline{\today}
\rightline{\tt hep-th/0701249}
\rightline{\tt MIT-CTP-3806}
\rightline{\tt DESY 07-007}
\rightline{\tt YITP-SB-07-3}
\begin{center}
\vskip 1.0cm
{\Large \bf {Analytic Solutions for Marginal Deformations   }}\\
\vskip 0.4cm
{\Large \bf {in Open String Field Theory}}
\vskip 1.0cm
{\large {Michael Kiermaier${}^1$, Yuji Okawa${}^2$, Leonardo Rastelli${}^3$,
and Barton Zwiebach${}^1$}}
\vskip 1.0cm
{\it {${}^1$ Center for Theoretical Physics}}\\
{\it {Massachusetts Institute of Technology}}\\
{\it {Cambridge, MA 02139, USA}}\\
mkiermai@mit.edu, zwiebach@mit.edu
\vskip 0.5cm
{\it {${}^2$ DESY Theory Group}}\\
{\it {Notkestrasse 85}}\\
{\it {22607 Hamburg, Germany}}\\
yuji.okawa@desy.de
\vskip 0.5cm
{\it {${}^3$ C.N. Yang Institute for Theoretical Physics}}\\
{\it {Stony Brook University}}\\
{\it {Stony Brook, NY 11794, USA}}\\
leonardo.rastelli@stonybrook.edu
\vskip 0.5cm
\vskip 1.0cm
{\bf Abstract}
\end{center}

\noindent
We develop a calculable analytic approach
to marginal deformations in open string field theory
using wedge states with operator insertions.
For marginal operators with regular operator products,
we construct analytic solutions to all orders in
the deformation parameter.
In particular, we  construct
an exact time-dependent solution
that describes
D-brane decay and
incorporates all $\alpha'$ corrections.
For marginal operators with singular operator products,
we construct  solutions by regularizing the singularity
and adding counterterms.
We explicitly carry out the procedure
to  third order in the deformation parameter.

\medskip

\end{titlepage}

\newpage

\baselineskip=17pt

\tableofcontents

\sectiono{Introduction}

Mapping the landscape of vacua is one
of the outstanding challenges in string theory.  A simpler version
of the problem is to characterize
the ``open string landscape,''
the set  of possible D-brane configurations in a fixed closed string background.
In recent years evidence has accumulated that
classical open string field theory (OSFT)
gives an accurate description of the open string landscape.
See \cite{Sen:2004nf, Taylor:2003gn, Taylor:2006ye} for reviews.
 Much of this
evidence is based on numerical work in level truncation,
and there remain many interesting questions. Is  the correspondence
between boundary
conformal field theories
and classical OSFT solutions
one-to-one? Is the OSFT action of a single D-brane
capable of describing configurations of multiple D-branes?
Answering these questions is likely to require
analytic tools. Important analytic progress was made by Schnabl~\cite{Schnabl:2005gv}.
He found the exact solution corresponding to the tachyon vacuum
by exploiting the simplifications
coming from
the clever gauge fixing condition
\be \label{schnablgaugeI}
B \Psi = 0 \, ,
\ee
where $B$ is the antighost zero mode in the conformal frame of the sliver.
Various aspects of Schnabl's construction
have been studied in \cite{Okawa:2006vm}--\cite{Erler:2006hw}.

In this paper
we describe new analytic solutions of OSFT
corresponding to exactly marginal deformations of the boundary
conformal field theory (CFT).
Previous work on exactly marginal deformations
in OSFT \cite{SZ} was based on solving the level-truncated
equations of motion in Siegel gauge. The level-truncated
string field was determined as a function of the
vacuum expectation value
of the exactly marginal mode
fixed to an arbitrary finite value.
Level truncation
lifts the flat direction, but it was seen that as the level
is increased the
flat direction is recovered with better and better accuracy.
Instead, our approach is to expand the solution as
$\Psi_\lambda = \sum_{n=1}^\infty \lambda^n \Psi^{(n)}  $,
where $\lambda$ parameterizes
the exact flat direction.
We solve the equation of motion recursively
to find an analytic expression for $\Psi^{(n)}$. Our results are exact in that
we are solving the full OSFT equation of motion, but they are
perturbative in $\lambda$; by contrast, the results
of \cite{SZ} are approximate since the equation of motion has been level-truncated,
but they are non-perturbative in the deformation parameter.

The perturbative approach of this paper has
certainly been attempted earlier
using the Siegel gauge.
Analytic work, however, is out of the question
because in the  Siegel gauge  the Riemann surfaces associated
with $\Psi^{(n)}$, with $n>2$, are very complicated.
The new insight that makes the problem
tractable is to use, as in \cite{Schnabl:2005gv}, the remarkable
properties of wedge states with insertions
\cite{Rastelli:2000iu, Rastelli:2001vb, Schnabl:2002gg}.

We find qualitatively different results,
according to whether the matter vertex operator $V$ that generates
the deformation has
regular or singular operator products.
Sections 2 and 3 of the paper
are devoted to the case of
regular operator products,
and the case of singular operator products is discussed in section 4.
A key technical point is the calculation
of the action of $B/L$, where
$L = \{ \, Q_B, B \, \}$, on products of string fields.

If $V$ has regular operator products,
the equation of motion
can be systematically solved in the Schnabl gauge (\ref{schnablgaugeI}).
The solution takes a strikingly compact form
given in the CFT language by (\ref{psingeometric}),
and its geometric picture is presented in Figure~1.
The solution $\Psi^{(n)}$ is made
of a wedge state with $n$ insertions
of $c V$ on its boundary. The relative separations of the boundary insertions
are specified by  $n-1$ moduli $ \{ t_i \}$, with $0 \leq t_i \leq 1$, which are to be integrated over.
Each modulus is
accompanied by an antighost line integral ${\cal B}$.
The explicit evaluation of $\Psi^{(n)}$
in the level expansion
is straightforward for
a specific choice of $V$.

In \S3.2 we apply this general result to the
  operator $V = e^{\frac{1}{\sqrt{\alpha'}}X^0} $
\cite{rolling}--\cite{Coletti:2005zj}.
  This deformation describes a time-dependent tachyon solution that starts at the perturbative
vacuum in the infinite past
and (if $\lambda < 0$)
 begins to roll toward
the non-perturbative vacuum.
The parameter $\lambda$
can be rescaled
by a shift of the origin of time, so the solutions are physically equivalent.
The time-dependent tachyon field  takes the form
\be \label{Tx0}
T(x^0) =  \lambda \, e^{\frac{1}{\sqrt{\alpha'}}x^0} + \sum_{n=2}^\infty  \lambda^n\,  \beta_n \, e^{\frac{1}{\sqrt{\alpha'}} n x^0} \, .
\ee
We derive a
closed-form integral expression
for the coefficients $\beta_n$
and evaluate them numerically.
We find that
the coefficients decay so rapidly as $n$ increases that
it is plausible that the solution is
absolutely convergent for any value of $x^0$.
Our exact result confirms
the surprising oscillatory behavior found in
the $p$-adic model~\cite{Moeller:2002vx} and
in level-truncation studies of OSFT~\cite{Moeller:2002vx, Coletti:2005zj}.
The tachyon (\ref{Tx0}) overshoots the non-perturbative vacuum and
oscillates with ever-growing amplitude. It has been argued that
a field redefinition to the variables of boundary SFT would
map this oscillating tachyon to a  tachyon field monotonically relaxing
to the non-perturbative vacuum \cite{Coletti:2005zj}.
 It would be very interesting
to calculate the pressure of our exact solution and check
whether it tends to zero in the infinite future, as would be expected from Sen's analysis
of tachyon matter \cite{Sen:2002in, Sen:2004nf}.

In \S3.3 we consider the
lightcone
vertex operator
$\partial X^+$,
another example of a marginal vertex operator
with regular operator products.
Following~\cite{Michishita:2005se}, we construct
the string field solution inspired by the Born-Infeld solution that
describes a fundamental string ending on a D-brane~\cite{Callan:1997kz}.
The lightcone direction $X^+$ is a linear combination
of the time direction and a direction normal to the brane,
and the vertex operator is dressed by
$A(k_i) \, e^{i k_i X^i}$ and integrated over
the momenta $k_i$ along the spatial directions on the brane.
The solution is not fully self-contained within
open string field theory: it requires sources,
which makes the analysis delicate.
Sources are also required in the
Born-Infeld
description of the solution.

If the operator product expansion (OPE)
of $V$ with itself is $V(z) V(0) \sim 1/z^2$,
the solution presented in Figure~1 is not well defined
because divergences arise
as the separations $t_i $ of the boundary insertions go to zero.
We study the required modifications in section~4.
An important example is the
Wilson-line
deformation $\partial X$.
We
regularize the divergences by imposing a cut-off in the integration
region of the moduli.
It turns out that  counterterms can be added to
obtain $\Psi^{(2)}$ that is finite and satisfies the equation of motion.
Surprisingly, the result necessarily
violates the gauge condition (\ref{schnablgaugeI})!
The naive solution
$\Psi^{(2)} = - \frac{B}{L} (\Psi^{(1)} * \Psi^{(1)} ) $
breaks down because the string field $\Psi^{(1)} * \Psi^{(1)}$
contains a component in the kernel of $L$.
This phenomenon is a peculiar quirk  of Schnabl gauge
that has no counterpart in Siegel gauge.
Due to this technical complication,
the construction of the higher $\Psi^{(n)}$ becomes quite cumbersome,
though still simpler than in Siegel gauge. We argue
that  for all $n$, appropriate counterterms can be added
to achieve a finite $\Psi^{(n)}$ that solves the equation of motion.
We discuss in detail the case of $\Psi^{(3)}$
and verify
the nontrivial cancellations
that must occur for the construction
to succeed. We leave it for future work to achieve simpler
closed-form expressions
for $\Psi^{(n)}$. Such expressions will be needed
to investigate  the nature of the perturbative series
in $\lambda$ and to make contact with the non-perturbative,
but approximate, level-truncation results of \cite{SZ}.
It will  also be interesting to understand  better
the relation between the conditions for
exact marginality of boundary CFT \cite{Recknagel:1998ih}
and the absence of obstructions in solving the equation of motion
of string field theory.
The technology developed in this paper
will be also useful
in open superstring field theory \cite{Berkovits:1995ab}.

\medskip

Independent work by M. Schnabl
on the subject of marginal deformations in string field theory
appears in \cite{martin}.

\vskip 1cm

\sectiono{The action of $B/L$}

\subsection{Solving the equation of motion in the Schnabl gauge}

For any matter
primary field $V$ of dimension one,
the state $\Psi^{(1)}$ corresponding to the operator $cV (0)$
is BRST closed:
\begin{equation}
Q_B \Psi^{(1)} = 0 \,.
\end{equation}
In the context of string field theory,
this implies that the linearized equation of motion
of string field theory is satisfied.
When the marginal deformation associated with $V$
is {\it exactly} marginal, we expect that a solution of the form
\begin{equation}
\Psi_\lambda = \sum_{n=1}^\infty \lambda^n \, \Psi^{(n)} \,,
\end{equation}
where $\lambda$ is a parameter, solves the nonlinear equation of motion
\begin{equation}
\label{eom}
Q_B \Psi_\lambda + \Psi_\lambda \ast \Psi_\lambda = 0 \,.
\end{equation}
The equation that determines $\Psi^{(n)}$ for $n > 1 $ is
\begin{equation}
Q_B \Psi^{(n)}  = \Phi^{(n)}
\quad \hbox{with} \quad
\Phi^{(n)} = -  \sum_ {k=1}^{n-1} \Psi^{(n-k)} * \Psi^{(k)} \,.
\label{PertEom}
\end{equation}
For this equation to be consistent, $\Phi^{(n)}$
must be BRST closed. This is easily shown
using the equations of motion at lower orders.
For example,
\be
Q_B \Phi^{(2)} =-Q_B \, ( \, \Psi^{(1)} * \Psi^{(1)} \, )
= - Q_B \Psi^{(1)} \ast \Psi^{(1)}  +  \Psi^{(1)} \ast Q_B\Psi^{(1)}
= 0
\ee
when $Q_B \Psi^{(1)} = 0 \,$.
It is crucial that $\Phi^{(n)}$
be BRST {\it exact} for all $n>1$,
or else we would encounter an obstruction
in solving the equations of motion.
No such obstruction is expected to arise if the matter operator $V$ is
exactly marginal,
so we can determine $\Psi^{(n)}$ recursively
by solving $Q_B \Psi^{(n)} = \Phi^{(n)} \,$.
This procedure is ambiguous as we can add any BRST-closed
term to $\Psi^{(n)}$, so we need to choose some prescription.
A traditional choice would be to work in Siegel gauge.
The solution $\Psi^{(n)}$ is then given by acting with
$b_0/L_0$ on $\Phi^{(n)}$.
In practice this is cumbersome
since the combination of star products
and operators $b_0/L_0$ in the Schwinger representation
generates complicated Riemann surfaces
in the CFT formulation.

Inspired by Schnabl's success in finding
an analytic solution for tachyon condensation,
it is natural to look for a solution $\Psi_\lambda$ in
the Schnabl gauge:
\be \label{schnablgauge}
B \Psi_\lambda = 0 \,  .
\ee
Our notation is
the same as  in \cite{Okawa:2006vm, RZ, ORZ}.
In particular the operators $B$ and $L$
are the zero modes of  the antighost and of
the energy-momentum tensor $T$, respectively,
in the conformal frame of the sliver,\footnote{
Using
reparameterizations,
as in \cite{ORZ}, it should be
straightforward to generalize the discussion to general projectors.
In this paper we restrict ourselves
to the simplest case of the sliver.}
\be
B \equiv \oint \frac{d \xi}{2 \pi i} \, \frac{f(\xi)}{f'(\xi)} \,
b(\xi)  \, , \quad
L \equiv \oint \frac{d \xi}{2 \pi i} \, \frac{f(\xi)}{f'(\xi)} \,
T(\xi)  \, ,\quad
f(\xi) \equiv \frac{2}{\pi} \, \arctan(\xi) \,  .
\ee
We define
$L^\pm \equiv L \pm L^\star $ and $B^\pm \equiv B \pm B^\star$,
where the superscript $\star$ indicates
BPZ conjugation, and we denote with subscripts $L$ and $R$
the left and right parts, respectively, of these operators.
Formally, a solution of (\ref{PertEom})
obeying (\ref{schnablgauge})
can be constructed as follows:
\begin{equation}
\label{formal}
\Psi^{(n) } = \frac{B}{L} \Phi^{(n)} \,  .
\end{equation}
This can also be written as
\begin{equation}
\Psi^{(n) } = \int_0^\infty dT \, B e^{-T L} \, \Phi^{(n)} \,,
\label{formal-solution}
\end{equation}
if the action of $e^{-T L}$ on $\Phi^{(n)}$ vanishes
in the limit $T \to \infty$.
It turns out that the action of $B/L$ on $\Phi^{(n)}$
is not always well defined.
As we discuss in detail in section 4,
if the matter primary field $V$ has a singular OPE with itself,
the formal solution breaks down and the required modification
necessarily violates the gauge condition (\ref{schnablgauge}).
On the other hand,
if operator products of the matter primary field are regular,
the formal solution is well defined,
as we will confirm later.
In the rest of this section,
we study the expression (\ref{formal-solution}) for $n=2$
in detail.

\subsection{Algebraic preliminaries}

We prepare for our work by reviewing and deriving some useful algebraic identities.
For further details and conventions the reader can refer to~\cite{RZ, ORZ}.

An important role will be played by the operator
$L - L^+_L$
and the antighost analog $B - B^+_L$.  These operators are derivations
of the star algebra.
This is seen by writing the first one, for example,
as a sum of two familiar derivations
in the following way:
\be
L - L^+_L = {1\over 2} L^-  + {1\over 2} (L^+_R+ L^+_L) - L^+_L =
{1\over 2} L^-  + {1\over 2} (L^+_R- L^+_L) = {1\over 2} (L^- + K) \,.
\ee
We therefore have
\begin{equation}
(L-L^+_L) \, (\phi_1 * \phi_2)
= (L-L^+_L) \, \phi_1 \ast \phi_2
+ \phi_1 \ast (L-L^+_L) \phi_2 \,.
\end{equation}
Noting that
$L^+_L \, (\phi_1 * \phi_2) = L^+_L \, \phi_1 \ast \phi_2$,
we find
\begin{eqnarray}
L (\phi_1 * \phi_2) & = & L \phi_1 * \phi_2 +
~ \phi_1 * (L-L^+_L)\,\phi_2 \, ,\label{Laction}\\
B (\phi_1 * \phi_2) & = & B \phi_1 * \phi_2 + (-1)^{\phi_1}
 \phi_1 * (B-B^+_L)\phi_2 \, . \label{Baction}
\end{eqnarray}
Here and in what follows, a string field in the exponent of $-1$
denotes its Grassmann property: it is $0$ mod $2$
for a Grassmann-even string field and $1$ mod $2$
for a Grassmann-odd string field.
{}From
(\ref{Laction}) and (\ref{Baction}) we immediately deduce formulas
for products of multiple string fields.
For $B$, for example, we have
\be
B (\phi_1 * \phi_2 * \ldots \phi_n)   =  (B \phi_1) *\ldots *\phi_n
+\sum_{m=2}^n  (-1)^{\sum_{k=1}^{m-1}\phi_k}
 ~\phi_1 * \ldots  *(B-B^+_L) \phi_m  *\ldots * \phi_n \,.
\label{Bactions}
\ee
Exponentiation of (\ref{Laction}) gives
\be \label{Lexpaction}
e^{- T L }  ( \phi_1 * \phi_2  ) = e^{-T L} \phi_1  * e^{-T (L-L^+_L)} \phi_2  \,.
\ee
{}From
the familiar commutators
\be
[L, L^+] =L^+ \, , \quad [B, L^+] = B^+ \, ,
\ee
we deduce
\be
[L , L^+_L ] = L^+_L \, , \quad [B, L^+_L ] = B^+_L \, .
\ee
See section 2 of \cite{RZ} for a careful analysis
of this type of manipulations.
We will need to reorder exponentials of the derivation
$L-L^+_L$.
We claim that
\begin{equation}
\label{ordering}
e^{-T (L-L^+_L)} = e^{(1-e^{-T}) L^+_L} \, e^{-T L} \,.
\end{equation}
The above is a particular case of
the Baker-Campbell-Hausdorff formula for a two-dimensional
Lie algebra with generators $x$ and $y$ and commutation relation
$[\, x , y \,] = y$.  In the adjoint representation we can write
\be
x = \begin{pmatrix} 0 & 1 \cr 0 & 1 \end{pmatrix}\,, \quad
y = \begin{pmatrix} -1 & 1 \cr -1 & 1 \end{pmatrix} \,.
\ee
It follows that as two-by-two matrices,  $x^2 = x$, $ xy= y$,  $yx =0$, and $y^2= 0$.
One then verifies that
\be
\label{cbh_2_dim}
e^{\alpha x + \beta y }
=  e^{{\beta\over \alpha} (e^\alpha -1) y}  \, e^{\alpha x}
\quad \hbox{when} \quad
[\, x, y \,] = y \,.
\ee
With $\alpha = -\beta = -T$, $x= L$, and $y = L^+_L$,
(\ref{cbh_2_dim}) reproduces (\ref{ordering}).

\subsection{The action of $B/L$ and its geometric interpretation}
\label{2.3}

We are now ready to solve the equation for $\Psi^{(2)}$.
The state $\Psi^{(1)}$ satisfies
\be \label{BL1}
Q_B \Psi^{(1)} = 0 \,, \qquad
B \Psi^{(1)} = 0 \,, \qquad
L \Psi^{(1)} = 0 \,.
\ee
We will use correlators in the sliver frame
to represent states made of wedge states and operator insertions.
The state $\Psi^{(1)}$ can be described as follows:
\begin{equation}
\label{psi1cft}
\langle \, \phi, \Psi^{(1)} \, \rangle
= \langle \, f \circ \phi (0) \, c V (1) \, \rangle_{{\cal W}_1} \,.
\end{equation}
Note that $cV$ is
a primary field of dimension zero
so that there is no associated conformal factor.
Here and in what follows we use $\phi$ to denote
a generic state in the Fock space
and $\phi (0)$ to denote its corresponding operator.
The surface ${\cal W}_\alpha$ is the one associated
with the wedge state $W_\alpha$ in the sliver conformal frame.
We use the doubling trick in calculating correlators.
We define the oriented straight lines $V^\pm_\alpha$ by
\begin{equation}
\begin{split}
& V^\pm_\alpha = \Bigl\{ \, z \, \Big| \,
{\rm Re} (z) = {}\pm \frac{1}{2} \, (1+\alpha) \, \Bigr\} \,, \\
& \hbox{orientation} \, :
{}\pm \frac{1}{2} \, (1+\alpha) -i \, \infty
\to {}\pm \frac{1}{2} \, (1+\alpha) +i \, \infty \,.
\end{split}
\end{equation}
The surface ${\cal W}_\alpha$ can be represented
as the region between $V^-_0$ and $V^+_{2 \alpha}$,
where $V^-_0$ and $V^+_{2 \alpha}$
are identified by translation.

A formal solution to the equation
$Q_B \Psi^{(2)} = - \Psi^{(1)} \ast \Psi^{(1)}$ is
\be
\Psi^{(2)}
= -  \int_0^\infty dT \, Be^{- T L}
\bigl[\, \Psi^{(1)} * \Psi^{(1)}\bigr] \,.
\ee
By construction, $B\Psi^{(2)} =0$. Using the identities
(\ref{Lexpaction}) and (\ref{Baction}), we have
\begin{equation}
\Psi^{(2)}
 = - \int_0^\infty dT \left[ \,
B \, e^{- T L} \, \Psi^{(1)}
\ast e^{- T (L-L^+_L)} \, \Psi^{(1)}
- e^{-T L} \, \Psi^{(1)}
\ast (B-B^+_L) \, e^{-T (L-L^+_L)} \, \Psi^{(1)} \, \right] \,.
\end{equation}
Because of the properties of $\Psi^{(1)}$ in (\ref{BL1}),
the first term vanishes and the second reduces to
\begin{equation}
\label{psi2_first_version}
\Psi^{(2)}
 = \int_0^\infty dT \, \Psi^{(1)}
\ast (B-B^+_L) \, e^{-T (L-L^+_L)} \, \Psi^{(1)} \,.
\end{equation}
We further use the identity (\ref{ordering})
together with $L\Psi^{(1)}=0$ to find
\be
\Psi^{(2)}= \int_0^\infty dT \,
\Psi^{(1)} \ast (B-B^+_L) \, e^{(1-e^{-T}) L^+_L} \, \Psi^{(1)}\,.
\ee
It follows from $[ \, B, L^+_L \,] = B^+_L$ that
$ [ \, B, \, g(L^+_L) \, ] = B^+_L \, g'(L^+_L)$
for any analytic function $g$.
Using this formula with $B\Psi^{(1)}=0 \,$, we find
\begin{equation}
\Psi^{(2)}=- \int_0^\infty dT \,
e^{-T} \, \Psi^{(1)} \ast e^{(1-e^{-T}) L^+_L} \,
B^+_L \Psi^{(1)} \,.
\end{equation}
Using the change of variables  $t= e^{-T}$, we obtain
the following final expression of $\Psi^{(2)}$:
\begin{equation}
\label{psi2t}
\Psi^{(2)} = \int_{0}^1 dt \, \Psi^{(1)} \ast e^{-(t-1) L^+_L} \,
(-B^+_L) \, \Psi^{(1)} \,.
\end{equation}

There is a simple geometric picture for $\Psi^{(2)}$.
Let us represent $\langle \, \phi, \Psi^{(2)} \, \rangle$
in the CFT formulation.
The exponential action of $L^+_L$ on a generic
string field $A$ can be written as
\be \label{expwedge}
e^{-\alpha L^+_L} A = e^{-\alpha L^+_L} ({\cal I} * A) = e^{-\alpha L^+_L } {\cal I} * A = W_\alpha * A \, .
\ee
Here we have recalled the familiar expression
of the wedge state
$W_\alpha =e^{-\frac{\alpha}{2} L^+} \mathcal{I}
= e^{-\alpha L^+_L } {\cal I}\,$ \cite{Schnabl:2005gv},
where  ${\cal I}$ is the identity string field.
We thus learn that $e^{-\alpha L^+_L}$ with $\alpha > 0$
creates a semi-infinite strip with a width of $\alpha$
in the sliver frame,
while $e^{-\alpha L^+_L}$ with $\alpha < 0$
deletes a semi-infinite strip with a width of $|\alpha|$.
The inner product $\langle \, \phi, \Psi^{(2)} \, \rangle$
is thus represented by a correlator
on ${\cal W}_{2-|t-1|}={\cal W}_{1+t}$.
In other words, the integrand in (\ref{psi2t})
is made of the wedge state $W_{1+t}$ with operator insertions.
The state $\phi$ is represented by the region
between $V^-_0$ and $V^+_0$
with the operator insertion $f \circ \phi (0)$ at the origin.
The left factor of $\Psi^{(1)}$  in (\ref{psi2t})
can be represented by the region between $V^+_0$ and $V^+_2$
with an insertion of $cV$ at $z=1$.
For $t=1$
the right factor of $\Psi^{(1)}$
can be represented by the region between $V^+_2$ and $V^+_4$
with an insertion of $cV$ at $z=2$.
For $0 < t < 1$,
the region is shifted
to the one between $V^+_{2-2 |t-1|}=V^+_{2t}$
and $V^+_{4-2 |t-1|}=V^+_{2+2t}$,
and the insertion of $cV$ is at $z=2-|t-1|=1+t$.
Finally, the operator $(-B_L^+)$ is represented
by an insertion of ${\cal B}$
\cite{ORZ} defined by
\begin{equation}
{\cal B} = \int \frac{dz}{2 \pi i} \, b(z) \,,
\end{equation}
where the contour of the integral can be taken to be
$-V^+_\alpha$ with $1 < \alpha < 1+2t$.
We thus have
\begin{equation}
\label{psi2_def}
\langle \, \phi, \Psi^{(2)} \, \rangle
= \int_0^1 dt \, \langle \, f \circ \phi (0) \,
c V (1) \, {\cal B} \, c V (1+t) \,
\rangle_{{\cal W}_{1+t}} \,.
\end{equation}
As $t\to 0$ the pair of $cV$'s collide,
and at $t=1$ they attain the maximum separation.

The state $\Psi^{(2)}$ should formally solve
the equation of motion by construction.
Let us examine the BRST transformation of $\Psi^{(2)}$ more carefully
based on the expression (\ref{psi2_def}).
The BRST operator in $\langle \, \phi, Q_B \Psi \, \rangle$
can be represented as an integral of the BRST current
on $V^+_{2(1+t)}-V^+_0$:\footnote{
To derive this we first use the relation
$\langle \, \phi, Q_B \Psi^{(2)} \, \rangle
= -(-1)^{\phi} \, \langle \, Q_B \phi,  \Psi^{(2)} \, \rangle$,
where $Q_B$ on the right-hand side
is an integral of
the BRST current $j_B$
over a contour that encircles the origin counterclockwise, with the operator
 $j_B$ placed to the left of $f\circ \phi(0)$ in the correlator.
Using the identification of the surface ${\cal W}_{1+t}$,
the contour can be deformed to $-V^+_{2(1+t)}+V^+_0$.
In the correlator, we move the BRST current  from
the left of $f \circ \phi (0)$
to the right of it. This cancels $(-1)^\phi$,
and the additional minus sign is canceled
by reversing the orientation of the contour.
}
\be
\langle \, \phi, Q_B \Psi^{(2)} \, \rangle  = \int_0^1 dt \, \Bigl\langle \, f \circ \phi (0) \,
\int_{\hskip-8pt {}_{-V^+_0 +V^+_{2(1+t)}}} \hskip-15pt {dz\over 2\pi i} \,
j_B(z) ~c V (1) \, {\cal B} \, c V (1+t) \,
\Bigr\rangle_{{\cal W}_{1+t}} \,,
\ee
where $j_B$ is the BRST current.
Since $cV$ is BRST closed,
the only nontrivial action of the BRST operator
is to change the insertion of
the antighost
to that of the energy-momentum tensor:
\begin{equation}
\langle \, \phi, Q_B \Psi^{(2)} \, \rangle
= {}- \int_0^1 dt \, \langle \, f \circ \phi (0) \,
c V (1) \, {\cal L} \, c V (1+t) \,
\rangle_{{\cal W}_{1+t}} \,,
\label{Q_B-Psi^(2)}
\end{equation}
where
\begin{equation}
{\cal L} = \int \frac{dz}{2 \pi i} \, T(z) \,,
\end{equation}
and the contour of the integral can be taken to be
$-V^+_\alpha$ with $1 < \alpha < 1+2t$.
The minus sign on the right-hand side of (\ref{Q_B-Psi^(2)})
is from anticommuting the BRST current with the left $cV$.
Since $\partial_t \, e^{-t L^+_L} = -L^+_L \, e^{-t L^+_L}$
and $-L^+_L$ corresponds
to ${\cal L}$
in the correlator,
an insertion of ${\cal L}$ is equivalent
to taking a derivative with respect to $t$ \cite{Okawa:2006vm}.
We thus find
\begin{equation} \label{taking}
\langle \, \phi, Q_B \Psi^{(2)} \, \rangle
= {}- \int_0^1 dt \, \frac{\partial}{\partial t} \,
\langle \, f \circ \phi (0) \, c V (1) \, c V (1+t) \,
\rangle_{{\cal W}_{1+t}} \,.
\end{equation}
The surface term from $t=1$ gives $- \Psi^{(1)} \ast \Psi^{(1)}$.
The equation of motion is therefore satisfied
if the surface term from $t=0$ vanishes.
The surface term from $t=0$ vanishes if
\begin{equation}
\lim_{t \to 0} \, cV (0) \, cV (t) = 0 \,.
\label{cV-condition}
\end{equation}
Therefore, $\Psi^{(2)}$ defined by (\ref{psi2_def}) does solve
the equation
$Q_B \Psi^{(2)} + \Psi^{(1)} \ast \Psi^{(1)} = 0$
when $V$ satisfies (\ref{cV-condition}).
Since $\Psi^{(1)} \ast \Psi^{(1)}$ is a finite state,
the equation guarantees that $Q_B \Psi^{(2)}$ is also finite.
However, it is still possible that $\Psi^{(2)}$
has a divergent term which is BRST closed.
The ghost part of $\Psi^{(2)}$ is finite
since it is given by an integral of $\psi_t$ over $t$ from $t=0$ to $t=1$,
where $\psi_n$  is the
key ingredient in the tachyon vacuum solution~\cite{Schnabl:2005gv}:
\begin{equation}
\langle \, \phi, \psi_n \, \rangle
= \langle \, f \circ \phi (0) \,
c (1) \, {\cal B} \, c (1+n) \,
\rangle_{{\cal W}_{1+n}} \,,
\label{psi_n}
\end{equation}
and the contour of the integral for ${\cal B}$
can be taken to be $-V_\alpha$ with $1 < \alpha < 2 \, n+1 \,$.
When the operator product of $V$ with itself is regular,
the condition (\ref{cV-condition}) is satisfied and
$\Psi^{(2)}$ itself is finite.
Note that $V(0) \, V(t)$ in the limit $t \to 0$
can be finite or can be vanishing.
We construct $\Psi^{(n)}$ for marginal operators
with regular operator products in the next section.
When the operator product of $V$ with itself is singular,
the formal solution $\Psi^{(2)}$ is not well defined.
We discuss this case in section 4.

\sectiono{Solutions for marginal operators
with regular operator products}\label{section_three}

In the previous section we  constructed a well-defined solution
to the equation $Q_B \Psi^{(2)} + \Psi^{(1)} \ast \Psi^{(1)} = 0$
when $V$ has a regular operator product.
In this section we generalize it to $\Psi^{(n)}$ for any $n$.
We then present the solution that corresponds to the decay of
an unstable D-brane in \S\ref{3.2}.
In \S\ref{3.3} we study marginal deformations
in the
lightcone
direction and discuss the application
to the solution that represents a string ending on a D-brane.

\subsection{Solution}\label{3.1}

Once we understand how $\Psi^{(2)}$ in the form of (\ref{psi2_def})
satisfies the equation of motion,
it is easy to construct $\Psi^{(n)}$
satisfying $Q_B \Psi^{(n)} = \Phi^{(n)}$.
It is given by
\begin{equation}\label{psingeometricxx}
\begin{split}
\langle \, \phi, \Psi^{(n)} \, \rangle
&  = \int_0^1 \hskip-3pt dt_1 \int_0^1 \hskip-3pt dt_2 \ldots
\int_0^1 \hskip-3pt dt_{n-1} \,
\langle \, f \circ \phi (0) \, c V (1) \, {\cal B} \,
c V (1+t_1) \, {\cal B} \, c V (1+t_1+t_2) \, \ldots \\
& \qquad \qquad \qquad \qquad \quad {}\times
{\cal B} \, c V (1+t_1+t_2+ \ldots +t_{n-1}) \,
\rangle_{{\cal W}_{1+t_1+t_2+ \ldots +t_{n-1}}}\, .
\end{split}
\end{equation}
Introducing the length parameters
\begin{equation}
\ell_i \equiv \sum_{k=1}^i t_k \,,
\end{equation}
the solution can be written more compactly as
\begin{equation}\label{psingeometric}
\boxed{ ~\langle \, \phi, \Psi^{(n)} \, \rangle
 = \int_0^1 dt_1 \int_0^1 dt_2 \ldots \int_0^1 dt_{n-1} \,
\Bigl\langle \, f \circ \phi (0) \, c V (1) \,
\prod_{i=1}^{n-1} \Bigl[ \, {\cal B} \, c V (1+ \ell_i) \,
\Bigr] \, \Bigr\rangle_{{\cal W}_{1+\ell_{n-1}}}\,. \,}
\end{equation}
See Figure~1.
The solution obeys the Schnabl gauge condition.
It is remarkably simple contrasted with the expression
one would obtain in Siegel gauge.

\begin{figure}
\centerline{\hbox{\epsfig{figure=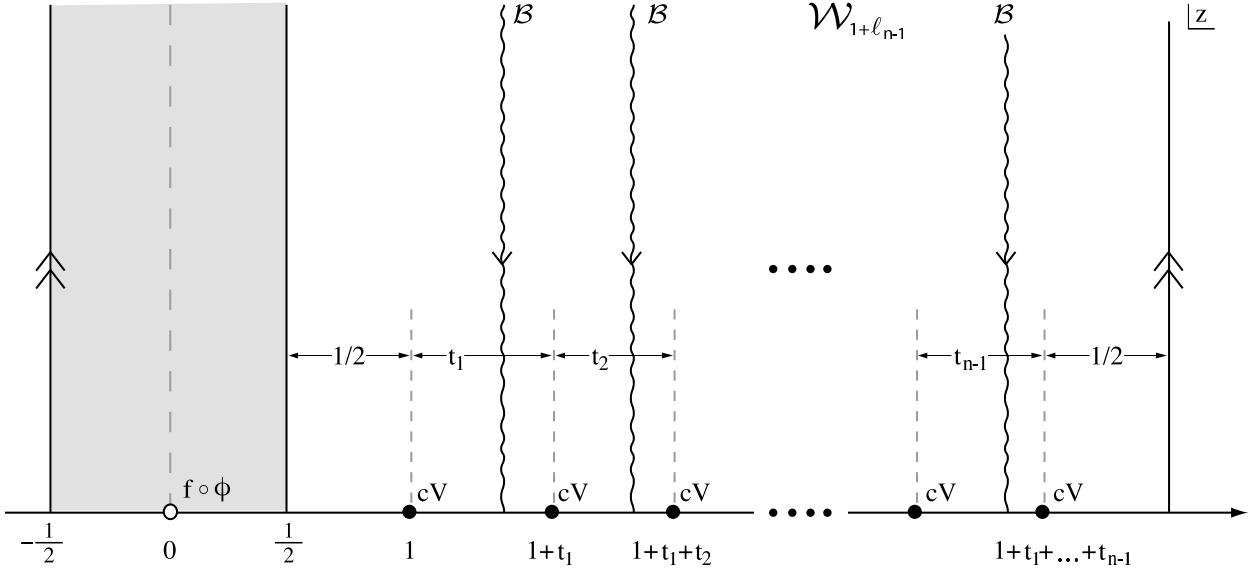, height=7.5cm}}}
\caption{The surface $\mathcal{W}_{1+\ell_{n-1}}$
with the operator insertions used to construct
the solution $\Psi^{(n)}$ given in (\ref{psingeometric}). The
parameters $t_1, t_2, \ldots,
t_{n-1}$ must all be integrated from
zero to one. The leftmost and rightmost vertical lines
with double arrows are identified.
}
\label{rz01f}
\end{figure}

\medskip
Let us now prove that the equation of motion is satisfied
for (\ref{psingeometric}).
It is straightforward to generalize the calculation
of $\langle \, \phi, Q_B \Psi^{(2)} \, \rangle$
in the previous section to that of
$\langle \, \phi, Q_B \Psi^{(n)} \, \rangle \,$.
The BRST operator in $\langle \, \phi, Q_B \Psi^{(n)} \, \rangle$
can be represented as an integral of the BRST current
on $V^+_{2(1+\ell_{n-1})}-V^+_0$.
Since $cV$ is BRST closed, the BRST operator acts
only on the insertions of ${\cal B}$'s:
\begin{equation}
\begin{split}
\langle \, \phi, Q_B \Psi^{(n)} \, \rangle
& = {}- \sum_{j=1}^{n-1} \int_0^1 dt_1 \int_0^1 dt_2 \ldots
\int_0^1 dt_{n-1} \, \Bigl\langle \, f \circ \phi (0) \, c V (1) \,
\prod_{i=1}^{j-1} \Bigl[ \, {\cal B} \,
c V (1+ \ell_i) \, \Bigr] \\
& \qquad \qquad {}\times
{\cal L} \, c V (1+ \ell_j) \,
\prod_{k=j+1}^{n-1} \Bigl[ \, {\cal B} \,
c V (1+ \ell_k) \, \Bigr] \,
\Bigr\rangle_{{\cal W}_{1+\ell_{n-1}}}\,.
\end{split}
\end{equation}
An insertion of ${\cal L}$
between $cV (1+\ell_{j-1})$ and $cV (1+\ell_j)$
corresponds to taking a derivative with respect to $t_j$.
When operator products of $V$ are regular, we have
\begin{equation}
\begin{split}
\langle \, \phi, Q_B \Psi^{(n)} \, \rangle
& = {}- \sum_{j=1}^{n-1} \int_0^1 dt_1 \int_0^1 dt_2 \ldots
\int_0^1 dt_{n-1} \, \partial_{t_j} \,
\Bigl\langle \, f \circ \phi (0) \, c V (1) \,
\prod_{i=1}^{j-1} \Bigl[ \, {\cal B} \,
c V (1+ \ell_i) \, \Bigr] \\
& \qquad \qquad {}\times c V (1+ \ell_j) \,
\prod_{k=j+1}^{n-1} \Bigl[ \, {\cal B} \,
c V (1+ \ell_k) \, \Bigr] \,
\Bigr\rangle_{{\cal W}_{1+\ell_{n-1}}} \\
& = {}- \sum_{j=1}^{n-1} \int_0^1 dt_1 \int_0^1 dt_2 \ldots
\int_0^1 dt_{j-1} \,
\int_0^1 dt_{j+1} \ldots \int_0^1 dt_{n-1} \,
\Bigl\langle \, f \circ \phi (0) \, c V (1) \\
& \qquad \qquad {}\times
\prod_{i=1}^{j-1} \Bigl[ \, {\cal B} \,
c V (1+ \ell_i) \, \Bigr] \, c V (1 + \ell_j) \,
\prod_{k=j+1}^{n-1} \Bigl[ \, {\cal B} \,
c V (1 + \ell_k) \, \Bigr] \,
\Bigr\rangle_{{\cal W}_{1 + \ell_{n-1}}} \biggr|_{t_j=1} \\
& = {}- \sum_{j=1}^{n-1} \, \langle \, \phi,
\Psi^{(j)} \ast \Psi^{(n-j)} \, \rangle \,.
\end{split}
\end{equation}
The equation of motion is thus satisfied.\footnote{
We assume that operator products of more than two $V$'s
are also regular
in order for the surface term from $t_j = 0$ to vanish.
This additional regularity condition was overlooked
in the first version of the paper on arXiv.
}

\bigskip
We can also derive this expression of $\Psi^{(n)}$ by
acting with $B/L$ on $\Phi^{(n)}$.
It is in fact interesting to see
how the region of the integrals over $t_1, t_2, \ldots, t_{n-1}$
is reproduced.
Let us demonstrate it taking the case of $\Psi^{(3)}$ as an example.
Using the Schwinger representation of $B/L$,
the expression (\ref{psi2_first_version}) for $\Psi^{(2)}$,
and the identities (\ref{Lexpaction})
and (\ref{Bactions}),
we have
\begin{equation}
\begin{split}
\Psi^{(3)}
& = - \int_0^\infty d T_2 \, B \, e^{-T_2 L} \,
[ \, \Psi^{(1)} \ast \Psi^{(2)} + \Psi^{(2)} \ast \Psi^{(1)} \, ] \\
& = - \int_0^\infty d T_2 \int_0^\infty dT_1 \, B \, e^{-T_2 L} \,
[ \, \Psi^{(1)} \ast \Psi^{(1)}
\ast (B-B^+_L) \, e^{-T_1 (L-L^+_L)} \, \Psi^{(1)} \\
& \qquad \qquad \qquad \qquad \qquad \qquad \qquad {}+ \Psi^{(1)}
\ast (B-B^+_L) \, e^{-T_1 (L-L^+_L)} \, \Psi^{(1)}
\ast \Psi^{(1)} \, ] \\
& = \int_0^\infty d T_1 \int_0^\infty \hskip-3pt d T_2 \,
[ \, \Psi^{(1)} \ast (B-B^+_L) \, e^{-T_2 (L-L^+_L)} \,
\Psi^{(1)} \ast (B-B^+_L) \, e^{-(T_1+ T_2) (L-L^+_L)} \,
\Psi^{(1)} \\
& \qquad \qquad \qquad \quad  {}+ \Psi^{(1)}
\ast (B-B^+_L) \, e^{-(T_1+T_2) (L-L^+_L)} \, \Psi^{(1)}
\ast (B-B^+_L) \, e^{-T_2 (L-L^+_L)} \, \Psi^{(1)} \, ] \,.
\end{split}
\end{equation}
By changing variables
as $\tau_1 = T_2$ and $\tau_2 = T_1+T_2$ for the first term
and as $\tau_2=T_2$ and $\tau_1 = T_1+T_2$ for the second term,
the two terms combine into
\be
\label{psi3combined}
\Psi^{(3)} = \int_0^\infty d\tau_1 \, \int_0^\infty d\tau_2 \,
\Psi^{(1)} \ast (B-B^+_L) \, e^{-\tau_1 (L-L^+_L)} \,
\Psi^{(1)} \ast (B-B^+_L) \, e^{-\tau_2 (L-L^+_L)} \,
\Psi^{(1)} \,.
\ee
The same manipulations we performed with $\Psi^{(2)}$ give
\begin{equation}
\begin{split}
\Psi^{(3)}
& = \int_0^1 dt_1 \, \int_0^1 dt_2 \,
\Psi^{(1)} \ast e^{-(t_1-1) L_L^+} (-B^+_L) \, \Psi^{(1)}
\ast e^{-(t_2-1) L_L^+} (-B^+_L) \, \Psi^{(1)}
\end{split}
\end{equation}
and the following expression in the CFT formulation:
\begin{equation}
\langle \, \phi, \Psi^{(3)} \, \rangle
 = \int_0^1 dt_1 \int_0^1 dt_2 \, \langle \, f \circ \phi (0) \,
c V (1) \, {\cal B} \, c V (1+t_1) \, {\cal B} \, c V (1+t_1+t_2) \,
\rangle_{{\cal W}_{1+t_1+t_2}}
\end{equation}
in agreement with (\ref{psingeometric}).
It is not difficult to use induction to prove that for all $n$ (\ref{psingeometric})  follows from the action of $B/L$ on $\Phi^{(n)}$.

\bigskip
We conclude the subsection
by writing other forms of the solution
that are suitable for explicit calculations.
We represent the surface ${\cal W}_\alpha$
as the region between
$V^-_2$ and $V^+_{2(\alpha-1)}$.
The operator $cV (1+\ell_{n-1})$ in (\ref{psingeometric})
is then mapped to $cV(-1)$.
We further transform
$\langle \, \phi, \Psi^{(n+1)} \, \rangle$
in the following way:
\begin{equation}\label{rearrange}
\begin{split}
\langle \, \phi, \Psi^{(n+1)} \, \rangle
&= \int_0^1 dt_1  \ldots \int_0^1 dt_{n} \,
\Bigl\langle c V (-1) \, f \circ \phi (0) \, c V (1) \,
\prod_{i=1}^{n-1} \Bigl[{\cal B} \, c V (1+ \ell_i) \,\Bigr]{\cal B} \,
 \Bigr\rangle_{{\cal W}_{1+\ell_{n}}} \\
&= \int_0^1 dt_1 \ldots \int_0^1 dt_{n} \,
\Bigl\langle c V (-1) \, f \circ \phi (0) \, c V (1) \,
\prod_{i=1}^{n-1} \Bigl[V (1+ \ell_i) \,\Bigr]{\cal B} \, \Bigr\rangle_{{\cal W}_{1+\ell_{n}}} \\
&= {}- \int_0^1 dt_1 \ldots \int_0^1 dt_{n} \, \frac{1}{2+\ell_n} \\
& \qquad \quad {}\times \Bigl\langle \,
\int_{V^+_{2 \ell_n}-V^-_2} \frac{dz}{2\pi i} \, z \, b(z) \,
\Bigl[ \, c V (-1) \, f \circ \phi (0) \, c V (1) \, \Bigr] \,
\prod_{i=1}^{n-1} \Bigl[ \, V (1+ \ell_i) \, \Bigr] \,
\Bigr\rangle_{{\cal W}_{1+\ell_{n}}}.
\end{split}
\end{equation}
First we recursively used the relation
${\cal B} \, c(z) \, {\cal B} = {\cal B} \, $,
which follows from $\{ \, {\cal B}, \, c(z) \, \} = 1$
and ${\cal B}^2 = 0 \,$.
In the last step, we used the identity
\begin{equation}
\int_{V^+_{2(\alpha-1)}-V^-_2}
\frac{dz}{2 \pi i} \, z \, b(z)
= (\alpha+1) \int_{V^+_{2(\alpha-1)}} \frac{dz}{2 \pi i} \, b(z)
\quad \hbox{on} \quad {\cal W}_\alpha \,.
\end{equation}
This follows from
\begin{equation}
\int_{V^-_2}
\frac{dz_-}{2 \pi i} \, z_- \, b(z_-)
= \int_{V^+_{2(\alpha-1)}}
\frac{dz_+}{2 \pi i} \,
\Bigl\{ \, z_+ - (\alpha+1) \, \Bigr\} \, b(z_+)
\quad \hbox{on} \quad {\cal W}_\alpha \,,
\end{equation}
where the coordinate $z_-$ for $V^-_2$
and the coordinate $z_+$ for $V^+_{2(\alpha-1)}$
are identified by $z_+ = z_- +\alpha+1 \,$.
The contour $V^+_{2 \ell_n}-V^-_2$ can be deformed
to encircle $c V (-1)$, $f \circ \phi (0)$, and $c V (1)$,
and we obtain
\begin{equation}\label{evaluateform}
\begin{split}
\langle \, \phi, \Psi^{(n+1)} \, \rangle
=\int_0^1 \hskip-4pt dt_1\ldots \int_0^1 \hskip-4pt dt_{n} \,
& \frac{1}{2+\ell_n} \,
\Bigl\langle ~ \Bigl\{ V (-1) \, f \circ \phi (0) \, c V (1)
 + c V (-1) \, f \circ \phi (0) \, V (1)  \\[1.0ex]
&\hskip-20pt + c V (-1) \,
\Bigl[ \, \oint \frac{dz}{2\pi i} \, z \, b(z) \,
f\circ \phi(0) \, \Bigr] \,
c V (1) \Bigr\} \prod_{i=1}^{n-1}V (1+ \ell_i) \,
\Bigl\rangle_{{\cal W}_{1+\ell_{n}}} \,,
\end{split}
\end{equation}
where the contour in the last line
encircles the origin counterclockwise.

When $\phi (0)$ factorizes into
a matter part $\phi_m (0)$ and a ghost part $\phi_g (0)$,
we can  use the matter-ghost factorization
of the correlator to give an alternative form of (\ref{psingeometric}):
\begin{equation}
\begin{split}
\langle \, \phi, \Psi^{(n)} \, \rangle
= \int_0^1 dt_1 \int_0^1 dt_2 \ldots \int_0^1 dt_{n-1} \,
& \Bigl\langle \, f \circ \phi_m (0) \,
\prod_{i=0}^{n-1} V (1+ \ell_i) \,
\Bigr\rangle_{{\cal W}_{1+\ell_{n-1}} ,\, m} \, \\
& {}\times
\Bigl\langle \, f \circ \phi_g (0) \, c(1) \,
{\cal B} \, c(1+\ell_{n-1}) \,
\Bigr\rangle_{{\cal W}_{1+\ell_{n-1}} ,\, g} \,,
\end{split}
\label{factorized-form}
\end{equation}
where $\ell_0 \equiv 0 \,$ and we denoted matter and ghost correlators by subscripts
$m$ and $g$,  respectively.
The ghost correlator in the above expression is
$\langle \, \phi_g, \psi_{\ell_{n-1}} \, \rangle$
in (\ref{psi_n}).  The algorithm for its calculation has been developed
in \cite{Schnabl:2005gv, ORZ}.

\subsection{Rolling tachyon marginal deformation to all orders}\label{3.2}

We can now apply the general solution (\ref{evaluateform}) to the special case of a marginal deformation corresponding to a rolling tachyon.
For this purpose we pick the operator
\begin{equation}
V(z, \bar z)
= e^{\frac{1}{\sqrt{\alpha'}}X^0(z, \bar z)}
\end{equation}
restricted to the boundary $z= \bar z = y$ of the upper-half plane $\mathbb{H}$, where we write it as\footnote{We use
the signature $(-, +, +, \ldots, +)$.
For a point
$z=\bar z = y$ on the boundary  of $\mathbb{H}$ we write $X^\mu(y)\equiv X^\mu(y, y)$.
The singular part of $X^\mu(y) X^\nu(y')$
is given by $-2\alpha' \eta^{\mu\nu} \ln |y-y'|$,
and the mode expansion for a Neumann
coordinate reads
$i \partial_y X^\mu(y)  = \sqrt{2\alpha'}  \sum_m  {\alpha^\mu_m \over y^{m +1}}$.
 The basic correlator
 is $\langle e^{ik\cdot X(y)}  e^{ik'\cdot X(y')} \rangle
= (2\pi)^D
\delta^{(D)}
(k+ k') |y-y'|^{2\alpha' k\cdot k'}$,
where $D$ is the spacetime dimension.
The operator
$e^{ik\cdot X(y)}$ has dimension $\alpha' k^2$ and transforms as
$f\circ e^{ik\cdot X(y)} = | {df\over dy}|^{\alpha' k^2} e^{ik\cdot X(f(y))}$.
We do not use the doubling trick
for the matter sector in \S\ref{3.2} and \S\ref{3.3}.
In these subsections,
$\partial X^\mu \equiv \partial_z X^\mu + \partial_{\bar{z}} X^\mu$
when $\mu$ is a direction along the D-brane and
$\partial X^\mu \equiv \partial_z X^\mu - \partial_{\bar{z}} X^\mu$
when $\mu$ is a direction transverse to the D-brane.}
\begin{equation}
V(y) = e^{\frac{1}{\sqrt{\alpha'}}X^0(y)}\,, \qquad   X^0 (y) \equiv X^0 (y, y) \,.
\end{equation}
The operator
$e^{ik\cdot X(y)}$ has dimension $\alpha' k^2$ and we can write
\be
V(y) = e^{ik\cdot X(y)}  \quad \hbox{with} \quad 
k^\mu=\frac{i}{\sqrt{\alpha'}} \bigl( 1, \vec 0\,\bigr)~
\to~\alpha' k^2 = 1\,,
\ee
showing that $V$ is
a matter primary field of dimension one.
We also have
\be
V(y) V(0) \sim   \, |y|^2   V(0)^2 \,,
\ee
and the matter operator
satisfies the requisite regularity condition.

We will also use exponential operators of $X^0$
with different exponents.
We thus record the following transformation law and
ordering results:
\be
f\circ e^{{1\over \sqrt{\alpha'}}\, nX^0(y)}  = \Bigl| {df\over dy}\Bigr|^{n^2}
e^{{1\over \sqrt{\alpha'}}\, nX^0(f(y))} \,,
\ee
\be
e^{{1\over \sqrt{\alpha'}}\, mX^0(y)}   e^{{1\over \sqrt{\alpha'}} \,nX^0(y')}
= |y-y'|^{2mn}  : e^{{1\over \sqrt{\alpha'}} \, mX^0(y)}   e^{{1\over \sqrt{\alpha'}} \,nX^0(y')} : \,.
\ee

\medskip
Physically,  deformation by $cV$ represents a rolling tachyon solution in which the state of the system at time $x^0=-\infty$ is the perturbative vacuum.
We set $\Psi^{(1)}$ to be
\be
\Psi^{(1)}  =   e^{\frac{1}{\sqrt{\alpha'}}X^0(0)} \,c_1|0\rangle
\ee
and calculate $\Psi^{(n)}$ with $n \ge 2$
which, by momentum conservation, must take the form
\be
\label{psi_n_tach}
\Psi^{(n)}  =   e^{\frac{1}{\sqrt{\alpha'}}\, nX^0(0)} \Bigl[ \beta_n \,c_1|0\rangle +
\ldots \Bigr] \,, \quad  n\geq 2 \,.
\ee
In the above we have separated out the tachyon component,
and higher-level fields
are indicated by dots.
The profile of the tachyon field $T$
is determined by the
coefficients $\beta_n$ that we aim to calculate:
\be
T(x^0) =  \lambda \, e^{\frac{1}{\sqrt{\alpha'}}x^0}  + \sum_{n=2}^\infty
\beta_n \, \lambda^n e^{\frac{1}{\sqrt{\alpha'}}\, nx^0}  \,.
\ee
Since the solution (for every component field) depends
on $\lambda$ and $x^0$ only through
the combination $\lambda e^{\frac{1}{\sqrt{\alpha'}}x^0}$,
a scaling of $\lambda$ can be absorbed by a shift of $x^0$. We can therefore focus on the case $\lambda=\mp 1$.
The
sign of $\lambda$ makes a physical difference. In our conventions the tachyon
vacuum lies at some $T<0$,
so $\lambda = -1$ corresponds to
the tachyon rolling
in the direction of the tachyon
vacuum, which
we are mostly interested in.  For $\lambda = +1$ the tachyon
begins to roll towards the unbounded region of the potential.
After setting $\lambda=\mp 1$, we write
\be
\label{tachy_profi}
T(x^0) = \mp \, e^{\frac{1}{\sqrt{\alpha'}}x^0}  + \sum_{n=2}^\infty
(\mp 1)^n
\beta_n \, e^{\frac{1}{\sqrt{\alpha'}}\, nx^0}  \,.
\ee

In order to extract the coefficients $\beta_n$ from the solution we introduce
test states $\phi_n$ and their BPZ duals:
\be
|\phi_n \rangle = e^{-\frac{1}{\sqrt{\alpha'}}\, nX^0(0)}\, c_0 c_1 |0\rangle \,,\qquad
\langle \phi_n |  =  \lim_{y\to \infty}  \langle 0 | c_{-1} c_0  e^{-\frac{1}{\sqrt{\alpha'}}\, nX^0(y)}\,{\,1\over |y|^{2n^2}} \,.
\ee
The state $\phi_n$ has dimension $n^2-1$. Using  (\ref{psi_n_tach}) we find
\be
\langle \phi_n,  \Psi^{(n)}\rangle = \langle \phi_n| \Psi^{(n)}\rangle =
\beta_n \cdot (\hbox{vol}) \,, \qquad   \hbox{vol} = (2\pi)^D
\delta^{(D)} (0) \,.
\ee
The spacetime volume (vol) always factors out,
so we will simply use
vol$\,=1$
in the following. We now  use (\ref{evaluateform}) to write
$\beta_{n+1} = \langle \phi_{n+1} , \Psi^{(n+1)} \rangle$ as
\begin{equation}\label{rollingPsin}
\begin{split}
\beta_{n+1} =\int_0^1 dt_1&\ldots \int_0^1 dt_{n} \, \frac{1}{2+\ell_n}
 \Bigl\langle   ~ \Bigl\{ e^{  {1\over  \sqrt{\alpha'}}  X^0(-1)  }  \, f \circ
 (\partial c) c  e^{-\frac{1}{\sqrt{\alpha'}}\, (n+1) X^0} (0) \, c
 e^{  {1\over  \sqrt{\alpha'}}  X^0}(1) \\[1.0ex]
 & \, + c e^{  {1\over  \sqrt{\alpha'}}  X^0  } (-1) \,
 f \circ (\partial c) c  e^{-\frac{1}{\sqrt{\alpha'}}\, (n+1) X^0} (0)
 \, e^{  {1\over  \sqrt{\alpha'}}  X^0(1)  }  \\[1.0ex]
&+ c e^{  {1\over  \sqrt{\alpha'}}  X^0  } (-1) \, f\circ c  e^{-\frac{1}{\sqrt{\alpha'}}\, (n+1) X^0} (0)  \, c e^{  {1\over  \sqrt{\alpha'}}  X^0  } (1)
 \Bigr\}
 \prod_{i=1}^{n-1} e^{  {1\over  \sqrt{\alpha'}}  X^0 (1+ \ell_i)  } \,
  \Bigl\rangle_{{\cal W}_{1+\ell_{n}}} .
\end{split}
\end{equation}
In the last term, due to the simple structure of $\phi_{n+1}$, the antighost line integral
acts as $b_0$ and simply removes the $c_0$ part of the state.
We must now evaluate the correlator
on the right-hand side.

This calculation
requires the map from
the surface
${\cal W}_{1+\ell_{n}}$ to the upper-half plane. We recall that
the surface ${\cal W}_0$ of unit width
can be mapped to the upper-half plane by the function
\begin{equation}\label{s32g}
g(z)=\frac{1}{2}\tan(\pi z)\,.
\end{equation}
Due to the periodicity $g(z+1)= g(z)$,
this map works independent of the position
of the surface ${\cal W}_0$
in the direction of the real axis.
Consequently, we merely need to rescale
${\cal W}_{1+\ell_{n}}$
to ${\cal W}_0$
by $z\rightarrow \frac{z}{2+\ell_n}$ and then
map it
to the upper-half plane
by $g(z)$.
The overall conformal transformation on the test states is therefore
the map $h$ given by
\begin{equation}\label{s32h}
h(\xi)\equiv g\Bigl(\frac{1}{2+\ell_n}\, f(\xi)\Bigr) \,.
\end{equation}
All other vertex operators are mapped with $g \bigl( {1\over 2+ \ell_n} \, z\bigr)$.  It is therefore natural to define
\begin{equation}\label{}
  g_i\equiv g\Bigl(\frac{1+ \ell_i}{2+\ell_n}\Bigr)\,, \qquad
  g'_i\equiv g'\Bigl(\frac{1+ \ell_i}{2+\ell_n}\Bigr)\,,
  \quad i = 0, 1, \ldots , n
  \,,  \quad\ell_0 \equiv 0\,.
\end{equation}
With these abbreviations, the correlator on the upper-half plane reads
\begin{equation}\label{}
\begin{split}
\beta_{n+1}
&=\int d^nt \,\frac{h'(0)^{(n+1)^2-1}}{2+\ell_n}
\biggl\langle
\Bigl\{\frac{g'_0}{2+\ell_n}\Bigl( e^{  {1\over  \sqrt{\alpha'}}  X^0(-g_0)  } \,
 (\del c) ce^{- {1\over  \sqrt{\alpha'}} (n+1)X^0}(0) \, c e^{ {1\over  \sqrt{\alpha'}} X^0}(g_0) \\[1.0ex]
 &\qquad+c e^{{1\over  \sqrt{\alpha'}}  X^0}(-g_0) \,
(\del c) c e^{-{1\over  \sqrt{\alpha'}}(n+1)X^0}(0) \,
e^{{1\over  \sqrt{\alpha'}} X^0(g_0)}\Bigr) \\
& \qquad
+ c e^{{1\over  \sqrt{\alpha'}} X^0} (-g_0) \,c e^{-{1\over  \sqrt{\alpha'}} (n+1)X^0} (0) \, c e^{{1\over  \sqrt{\alpha'}}X^0} (g_0)\Bigr\}
\,\prod_{i=1}^{n-1}\frac{g'_i}{2+\ell_n} e^{ {1\over  \sqrt{\alpha'}} X^0(g_i)} \, \, \biggr\rangle_{\mathbb{H}}\,,
\end{split}
\end{equation}
where
$h'(0)=\frac{1}{2+\ell_n}$ and
we have defined
$  \int d^nt\equiv\int_0^1 dt_1 \ldots \int_0^1 dt_{n} $.
We can now factor this into
matter and ghost correlators:
\begin{equation}\label{}
\begin{split}
\beta_{n+1}
=\int d^nt \,& (2+\ell_n)^{-(n+1)^2}\Bigl\langle e^{{1\over  \sqrt{\alpha'}} X^0(-g_0)}
 \, e^{-{1\over  \sqrt{\alpha'}}(n+1)X^0(0)} \, e^{{1\over  \sqrt{\alpha'}}  X^0(g_0)} \,
  \prod_{i=1}^{n-1}
\frac{g'_i}{2+\ell_n} e^{{1\over  \sqrt{\alpha'}}X^0(g_i)}  \, \Bigr\rangle_{m}
\\
& \times
\Bigl\langle \frac{g'_0}{2+\ell_n}\,\Bigl( (\del c) c(0) \, c(g_0)
+  c(-g_0) \, (\del c) c(0) \Bigr)
+ c(-g_0) \, c (0) \, c(g_0)
\Bigr\rangle_{g}.
\end{split}
\end{equation}
The ghost correlator can be evaluated  using
$\bigl\langle c(-z)c(0)c(z)\bigr\rangle_{g}=-2z^3$ and
$\bigl\langle \del c \, c(0) \, c(z)\bigr\rangle_{g}=z^2$. Using also
$-g_0=g_n$ and $g_0'=g_n'$, we find
\begin{equation}\label{}
\beta_{n+1}=2\int d^nt \, (2+\ell_n)^{-n(n+3)}
\Bigl(\frac{g'_0}{2+\ell_n} - g_0\Bigr)
\frac{g_0^2}{{g'_0}^2}
\prod_{i=0}^{n}
\Bigl[g'_i\Bigr]
\Bigl\langle \, e^{-{1\over  \sqrt{\alpha'}}(n+1)X^0(0)} \,
 \prod_{i=0}^{n} e^{{1\over  \sqrt{\alpha'}} X(g_i)} \,\, \Bigr\rangle_{m}\,.
\end{equation}
Evaluating the matter correlator,
we obtain our final result for the coefficients of the
rolling tachyon solution:
\begin{equation}\label{s3TachyCorr}
\boxed{
\beta_{n+1}=2\int d^nt \, (2+\ell_n)^{-n(n+3)}\Bigl(\frac{g'_0}{2+\ell_n} - g_0\Bigr)\frac{g_0^2}{{g'_0}^2}
\Bigl[\prod_{i=0}^{n}\frac{g'_i}{g_i^{2(n+1)}}\Bigr]\prod_{0\leq i<j\leq n}\bigl(g_i-g_j\bigr)^2. }
\end{equation}
Another way to derive (\ref{s3TachyCorr})
is to use (\ref{factorized-form}).
The ghost correlator,
which gives the tachyon coefficient of $\psi_{\ell_n}$,
has been calculated in \cite{Schnabl:2005gv, ORZ}:
\begin{equation}
\begin{split}
\langle \, f \circ (\partial c) c(0) \, c(1) \,
{\cal B} \, c(1+\ell_n) \, \rangle_{{\cal W}_{1+\ell_n},g}
& = \frac{2+\ell_n}{\pi} \,
\left[ \, 1-\frac{2+\ell_n}{2 \pi} \,
\sin \frac{2 \pi}{2+\ell_n} \, \right]
\sin^2 \frac{\pi}{2+\ell_n} \\
& = 2 \, (2+\ell_n) \, \frac{g_0^2}{g'_0} \,
\left( 1-\frac{(2+\ell_n) \, g_0}{g'_0} \right) \,.
\end{split}
\end{equation}
The calculation of the matter correlator is straightforward:
\begin{equation}
\begin{split}
& \Bigl\langle \,
f \circ e^{-\frac{1}{\sqrt{\alpha'}} \, (n+1) X^0 (0)} \,
\prod_{i=0}^n \, e^{\frac{1}{\sqrt{\alpha'}} \, X^0 (1+\ell_i)} \,
\big\rangle_{{\cal W}_{1+\ell_n} ,\, m} \\
& = \left( \frac{2}{\pi} \right)^{(n+1)^2}
\biggl[ \, \prod_{i=0}^n \,
\frac{(2+\ell_n)^{-2(n+1)}}{\pi^{-2(n+1)}} \,
\sin^{-2(n+1)} \frac{\pi (1+\ell_i)}{2+\ell_n} \, \biggr] \,
\prod_{0 \le i < j \le n} \, \frac{(2+\ell_n)^2}{\pi^2} \,
\sin^2 \frac{\pi (\ell_i-\ell_j)}{2+\ell_n} \\
& = (2+\ell_n)^{-(n+1)(n+2)}
\biggl[ \, \prod_{i=0}^n \,
\frac{g'_i}{g_i^{2(n+1)}} \, \biggr] \,
\prod_{0 \le i < j \le n} \, (g_i-g_j)^2 \,.
\end{split}
\end{equation}
It is easy to see that (\ref{s3TachyCorr}) is reproduced.

The integrand in (\ref{s3TachyCorr}) is manifestly positive since $g'(z) >0$  and $\frac{g'_0}{2+\ell_n} - g_0>0$. It follows that
all $\beta_{n+1}$ coefficients
are positive. For $n=1$ we find
\be
\label{}
\beta_2=8\int_0^1 dt \, \frac{\frac{g'_0}{2+t} - g_0}{(2+t)^{4} g_0^{4}}
=8\int_0^1 \hskip-4pt dt \, \biggl(\frac{2 \cot\bigl(\frac{\pi}{2+t}\bigr)}{2+t}\biggr)^4\Bigl(\frac{\pi}{2(2+t)\cos^2\bigl(\frac{\pi}{2+t}\bigr)} - \frac{1}{2}\tan\Bigl(\frac{\pi}{2+t}\Bigr)\Bigr)\,.
\ee
Surprisingly, analytic evaluation of the integral is possible using {\em Mathematica}:
\be
\beta_2 =\frac{64}{243\sqrt{3}}\,.
\ee
This coefficient is the same
as that of the Siegel-gauge solution~\cite{Coletti:2005zj}.
For $n=2$ the final integral
can be evaluated numerically:
\be
\beta_3= 8\int_0^1 dt_1\int_0^1dt_2 \, \frac{\Bigl(\frac{g'_0}{2+t_1+t_2} - g_0\Bigr)
g'_1\bigl(g_0^2-g_1^2\bigr)^2}{(2+t_1+t_2)^{10}g_0^8g_1^6} \simeq 2.14766\cdot 10^{-3} \,.
\ee
The results for the first few $\beta_n$ are summarized in
Table \ref{tab:TachyCoeff}.
The resulting tachyon profile (\ref{tachy_profi}) takes the form
\begin{equation}
\begin{split}
  T(x^0) =&\mp e^{\frac{1}{\sqrt{\alpha'}}\,x^0}
  +~0.15206 \,~ e^{\frac{1}{\sqrt{\alpha'}}\,2x^0}
~ ~ ~\,\,\mp 2.148\cdot 10^{-3} \, e^{\frac{1}{\sqrt{\alpha'}}\,3x^0}\\[0.5ex]
 &\qquad \qquad +2.619\cdot 10^{-6} \, e^{\frac{1}{\sqrt{\alpha'}}\,4x^0}
 \, \, \mp 2.791\cdot 10^{-10} \, e^{\frac{1}{\sqrt{\alpha'}}\,5x^0}\\[0.5ex]
  &\qquad\qquad +2.801\cdot 10^{-15} \, e^{\frac{1}{\sqrt{\alpha'}}\,6x^0}
  \mp 2.729\cdot 10^{-21} \, e^{\frac{1}{\sqrt{\alpha'}}\,7x^0}
  +\dots
\end{split}
\end{equation}
The top sign gives us the physical solution:
the tachyon rolls towards the tachyon vacuum,
overshoots it,
and then begins to
develop
larger and larger oscillations.
The coefficients in the solution decrease so rapidly that
the series seems to be absolutely convergent
for any value of ${x^0\over \sqrt{\alpha'}}$.  Indeed,  the
$n$-th term $T_n$  in the above series appears to take the form
\be
|T_n| \sim
2.7\cdot 10^{-{{1\over 2} n(n-1)}} \, e^{\frac{1}{\sqrt{\alpha'}}\, nx^0}\,.
\ee
One then finds that the ratio of consecutive coefficients is
\be
\Bigl| {T_{n+1}\over T_n} \Bigr| \sim  10^{-n} e^{\frac{1}{\sqrt{\alpha'}}\, x^0}
\simeq e^{-2.303 \,n} e^{\frac{1}{\sqrt{\alpha'}}\, x^0} \,.
\ee
For any value of ${x^0\over \sqrt{\alpha'}}$
the ratio becomes smaller than one for sufficiently large $n$,
suggesting
absolute convergence.  It would be useful to do
analytic estimates of $\beta_n$
using (\ref{s3TachyCorr}) to confirm the above speculation.

\begin{table}[t]
\begin{center}
\renewcommand{\arraystretch}{1.65}
\begin{tabular}{|c|c|}
\hline
$n$ & $\beta_{n}$\\
\hline\hline
2 & $\frac{64}{243\sqrt{3}}\approx 0.152059$\\
\hline
3 & $2.14766\cdot 10^{-3}$\\
\hline
4 & $2.61925\cdot 10^{-6}$\\
\hline
5 & $2.79123\cdot 10^{-10}$\\
\hline
6 & $2.80109\cdot 10^{-15}$\\
\hline
7 & $2.72865\cdot 10^{-21}$\\
\hline
\end{tabular}
\end{center}
\caption{Numerical values of the rolling tachyon profile coefficients.}
\label{tab:TachyCoeff}
\end{table}

It is interesting to compare
the results with those
of the $p$-adic model
\cite{Moeller:2002vx}. The relevant solution is discussed in \S4.2.2 of that paper and has the same qualitative behavior as the
solution presented here: the tachyon rolls towards the minimum, overshoots it,
and then
develops ever-growing
oscillations.  The solution is of the form
\be
\phi(t) = 1 - \sum_{n=1}^\infty  a_n \, e^{\sqrt{2} nt}\,, \quad  a_1 = 1\,.
\ee
The coefficients $a_n$ can be calculated exactly with a simple recursion and fall off
very rapidly, but an analytic expression for their large $n$ behavior is not known.
A fit of the values of
$a_n$ for $n=3, \ldots , 13$ gives
$\ln a_n \simeq  -0.1625  + 1.506 \,n  -1.389 \, n^2$.
(A
fit with an $n^3$ term returns a very small coefficient for this
term.)
The fit implies that the ratio of two consecutive terms in the solution~is
\be
\Bigl| {a_{n+1}\over a_n} \Bigr| e^{\sqrt{2}\, t}\sim e^{-2.778\, n  + 0.117} e^{\sqrt{2}\, t}\simeq  1.125 \cdot 16^{-n} e^{\sqrt{2}\, t}\,.
\ee
This result suggests that the $p$-adic rolling solution is also absolutely convergent.

A low-level
solution of the string theory rolling tachyon in Siegel gauge was also obtained
in~\cite{Moeller:2002vx},  where significant similarities with the $p$-adic solution were noted.
The higher-level Siegel gauge analysis of the rolling tachyon
in~\cite{Coletti:2005zj} confirmed the
earlier analysis and added much confidence to the validity of the oscillatory solution.
We believe that the exact analytic solution presented here has settled
the issue convincingly.

\subsection{Lightcone-like  deformations}\label{3.3}

Another simple example of a marginal operator with
regular operator products
is provided
by the lightcone-like operator
\be
V(z) =  \frac{i}{\sqrt{2\alpha'}} \, \del X^+ \,,
\ee
as usual, inserted at $z=\bar z = y$.
Here $X^+=\frac{1}{\sqrt{2}}(X^0+X^1)$
is a lightcone
coordinate. (We could have also chosen
$X^- = \frac{1}{\sqrt{2}}(X^0 - X^1)$.)
The OPE of $V$ with itself is regular:
$\lim_{z\to 0} V(z) V(0) = V(0)^2$.  The operator is dimension one and
$cV$ is BRST closed. We can
construct a solution
using the above $V(z)$ and our general result (\ref{evaluateform}).
If we consider some D$p$-brane
with $p < D-1$,
we can
choose $x^1$ to be a direction normal to the brane and
the above matter deformation
corresponds
to giving constant expectation values
to the time component of the gauge field on the brane and to the scalar field on the
brane that represents
the position
of the brane.

To make the analysis a bit more nontrivial we consider the discussion of
Michishita~\cite{Michishita:2005se}
on
the Callan-Maldacena
solution~\cite{Callan:1997kz} for a string ending on a brane in the framework of OSFT.
We
choose
\begin{equation}
  V(y)=\int dk_i \, A(k_i) \, \frac{i}{\sqrt{2\alpha'}} \, \del X^+e^{ik_iX^i}(y)\,,
\end{equation}
where $X^i$'s are the spatial directions on the brane.
This operator has
regular operator products:
the exponentials $e^{ik_iX^i}(y)$
give positive powers of distances
since
$k_i$ is spacelike.
The operator
$c\del X^+e^{ik_iX^i}$,
however, has dimension
$ \alpha' k^2$,  so unless $k_i=0$ it is not BRST closed
and the expression in~(\ref{evaluateform}) does not provide a solution.
But it is not too far from a solution: if one chooses $A(k)\sim 1/k^2$, the action of $Q_B$
on $cV$
gives
a delta function in position space.

We thus take
$ \Psi^{(1)}_A =  V(0) c_1 |0\rangle$  and, following~\cite{Michishita:2005se}, take its failure to be annihilated by
$Q_B$ to define the source term $J^{(1)}$ that hopefully would arise independently in a complete theory:  $Q_B \Psi_A^{(1)} = J^{(1)}$.
We can then calculate
$\Psi_A^{(2)}$ which satisfies
$Q_B \Psi_A^{(2)} + \Psi_A^{(1)} * \Psi_A^{(1)} = J^{(2)}$
for some $J^{(2)}$.
While  $BJ^{(1)}\neq0$,
we demand $BJ^{(n)}=0$ for $n\geq 2$
following the approach of~\cite{Michishita:2005se}
in the Siegel-gauge case.
Acting with $B$ on the above equation for $\Psi^{(2)}_A$,
we find
\be
\label{new_op_e}
L \Psi_A^{(2)} + B(\Psi_A^{(1)} * \Psi_A^{(1)})=0 \quad \to \quad
\Psi_A^{(2)} = -{B \over L}(\Psi_A^{(1)} * \Psi_A^{(1)})\,.
\ee
\medskip
Acting with $Q_B$ on the solution, one confirms that
\be
Q_B\Psi_A^{(2)} = - \Psi_A^{(1)} * \Psi_A^{(1)}
+ {B \over L}\Bigl( J^{(1)} * \Psi_A^{(1)} -\Psi_A^{(1)} *J^{(1)}\Bigr)
\ee
so that the source term $J^{(2)}$ is indeed annihilated by $B$.

In calculating $\Psi_A^{(2)}$ in~(\ref{new_op_e})
with $L \Psi^{(1)}_A \not= 0$,
we need to generalize our results in \S\ref{2.3} and
find the action of
$B/L$ on a string field product $\chi\ast\chi'$ where $\chi$ and $\chi'$ are not annihilated by $L$ but instead satisfy
\be \label{}
B \chi =B \chi'= 0 \,, \qquad
L \chi =l_\chi \, \chi \,, \qquad
L {\chi'} =l_{\chi'} \, \chi' \,.
\ee
The steps leading to~(\ref{psi2t}) can be carried out analogously
for this case
with extra factors depending on $l_\chi$ and $l_{\chi'}$:
\begin{equation}
\frac{B}{L}(\chi\ast\chi')
= (-1)^\chi
\int_{0}^1 dt \,t^{(l_\chi+l_{\chi'})} \, \chi \ast e^{-(t-1) L^+_L} \,
(-B^+_L) \, \chi' \,.
\end{equation}
To construct $\Psi^{(2)}_A$, we need to express states of the type $\frac{B}{L}(\chi\ast\chi')$ as CFT correlators. As $\chi$ and $\chi'$ are
primary fields
of nonvanishing dimension,
there are
extra conformal factors in the
sliver-frame expression
for these states. Defining a shift function $s_l(z)=z+l$, we can express the  generalization of~(\ref{psi2_def}) that accounts for these extra factors as
\begin{equation}
\begin{split}
\langle \, \phi, \frac{B}{L}(\chi\ast\chi') \, \rangle
&= (-1)^\chi
\int_0^1 dt \,t^{(l_\chi+l_{\chi'})} \, \langle \, f \circ \phi (0) \,\,\,
s_1\circ f\circ\chi(0) \,\, {\cal B} \,\,\, s_{1+t}\circ f\circ\chi'(0) \,
\rangle_{{\cal W}_{1+t}} \\
&= (-1)^\chi
\int_0^1 dt \,\bigl(tf'(0)\bigr)^{(l_\chi+l_{\chi'})} \, \langle \, f \circ \phi (0) \,\,
\chi(1) \,\, {\cal B} \,\, \chi'(1+t) \,
\rangle_{{\cal W}_{1+t}} \,.
\end{split}
\end{equation}
Here we have explicitly carried out the conformal maps of $\chi$ and $\chi'$ to the sliver frame and used $s'_l(z)=1$. It is now
straightforward
to carry out the construction of $\Psi_A^{(2)}$
by generalizing~(\ref{evaluateform}).
This yields
\begin{equation}\label{}
\begin{split}
\langle\phi, \Psi^{(2)}_A\rangle=\int dk_idk'_iA(k_i)A(k'_i)&\int_0^1 dt \, \frac{-\bigl(tf'(0)\bigr)^{\alpha'(k^2+k'^2)}}{(2+t)2\alpha'}
 \Bigl\langle   ~ \Bigl\{ \del X^+e^{ik_iX^i} (-1) \, f \circ \phi (0) \, c \del X^+e^{ik'_iX^i} (1) \\
&+ c \del X^+e^{ik_iX^i} (-1) \, \Bigl[\oint \frac{dz}{2\pi i}zb(z)f\circ \phi(0)\Bigr] \, c \del X^+e^{ik'_iX^i} (1)  \\
& \, + c \del X^+e^{ik'_iX^i} (-1) \, f \circ \phi (0) \, \del X^+e^{ik_iX^i} (1)\Bigr\} \, \Bigl\rangle_{{\cal W}_{1+t}} .
\end{split}
\end{equation}
To obtain
a Fock-space expression
of $\Psi_A^{(2)}$, we follow the same steps leading to (5.50) of~\cite{ORZ}. The map we need to perform on the correlator is $I\circ g$, so the total map on the test state $\phi$ is $I\circ h$. Here we have used $g$ and $h$
defined in~(\ref{s32g}) and~(\ref{s32h}),
and $I(z)=-\frac{1}{z}$.
Let us further define
\begin{equation}
  \hat{B}=\oint\frac{dz}{2\pi i}\>\frac{g^{-1}(z)}{(g^{-1})'(z)}b(z).
\end{equation}
Then we can start by mapping  the correlator to the upper-half plane through $g$. Again, we will suppress all arguments of $g$ and abbreviate
\begin{equation}\label{}
\begin{split}
  g\equiv g\Bigl(\frac{1}{2+t}\Bigr)
  =-g\Bigl(-\frac{1}{2+t}\Bigr) \,,
  \qquad\qquad g'\equiv g'\Bigl(\frac{1}{2+t}\Bigr)=g'\Bigl(-\frac{1}{2+t}\Bigr).
\end{split}
\end{equation}
We find
\begin{equation}\label{Psi2UHP}
\begin{split}
\langle\phi, \Psi^{(2)}_A\rangle=&\int dk_idk'_iA(k_i)A(k'_i)\int_0^1 dt \, \frac{-1}{(2+t)2\alpha'}\Bigl(\frac{t f'(0) g'}{2+t}\Bigr)^{\alpha'(k^2+k'^2)}\\
 & \times
\Bigl\langle \frac{g'}{2+t} \Bigl\{\del X^+e^{ik_iX^i} (-g) \, h \circ \phi (0) \, c \del X^+e^{ik'_iX^i} (g) \\
&\qquad + c \del X^+e^{ik'_iX^i} (-g) \, h \circ \phi (0) \, \del X^+e^{ik_iX^i} (g)\Bigr\}\\
&\qquad +
c \del X^+e^{ik_iX^i} (-g) \, \Bigl[\hat{B} \, h\circ \phi(0)\Bigr] \, c \del X^+e^{ik'_iX^i} (g) \, \Bigr\rangle_{\mathbb{H}} .
\end{split}
\end{equation}
Here we used the fact
 that the operator $c \del X^+e^{ik_iX^i}$ has conformal dimension $\alpha' k^2$. We notice that the two terms in
parenthesis
can be
transformed
into each other through the map $g\rightarrow-g$. Therefore, we can drop one of them and simply take the $g$-even part of the other. We can now perform the remaining transformation with $I$ to obtain an operator expression for $\Psi_A^{(2)}$:
\begin{equation}\label{Psi2operator}
\begin{split}
\Psi^{(2)}_A&=\int dk_idk'_i \, A(k_i)A(k'_i)\int_0^1 dt \, \frac{-1}{(2+t)2\alpha'}\Bigl(\frac{t f'(0) g'}{(2+t)g^2}\Bigr)^{\alpha'(k^2+k'^2)}\\
 &\quad \times
U^\star_{h}\biggl[\Bigl\{\frac{2g'}{(2+t)g^2} \, \del X^+e^{ik_iX^i} \bigl(-1/g\bigr) \, c \del X^+e^{ik'_iX^i} \bigl(1/g\bigr)
\Bigr\}_{g\textrm{-even}} \\
 &\qquad\qquad\qquad+ \hat{B}^\star \, c \del X^+e^{ik_iX^i} \bigl(-1/g\bigr) \, c \del X^+e^{ik'_iX^i} \bigl(1/g\bigr) \, \biggr]\ket{0}\\
 & \equiv \int dk_idk'_i \, A(k_i)A(k'_i)\, \Psi^{(2)}_{k,k'} \,.
\end{split}
\end{equation}
We would now like to determine the level expansion of $\Psi^{(2)}_A$,
or equivalently,
of its momentum decomposition $\Psi^{(2)}_{k,k'}$. We can either attempt a direct level expansion of the operator result~(\ref{Psi2operator})
or use
the test state formalism that we carried out
in \S\ref{3.2}. It is straightforward to carry out the first method for the case of vanishing momentum $k=k'=0$, so we will start with this approach. We will then use the test state method to find the level expansion with full momentum dependence.

Let us start by
the level expansion of $\Psi^{(2)}_{k,k'}$ in~(\ref{Psi2operator}).
We use the results in~\S6.1 of~\cite{ORZ}
to obtain the following useful expansions:
\begin{equation}
\begin{split}
\hat{B}^\star=b_0+\frac{8}{3}b_{-2}+\dots \,,
\qquad\qquad U^\star_h=(2+t)^{-L_0}+\dots
\end{split}
\end{equation}
Here the dots denote
higher-level corrections.
We notice that
self-contractions of $\del X^+$ vanish
as $\eta^{++}=0$. We end up with the following mode expansions for the matter and ghost fields:
\begin{equation}
\begin{split}
  -\frac{1}{2\alpha'}\,
  \del X^+(-1/g)\del X^+(1/g)\ket{0}&=
  \sum_{i<0,j<0}
  (-1)^{i+1}
  (\alpha^+_i\alpha^+_j)g^{i+j+2}\ket{0}\,,\\
  c(\pm1/g)=
  \sum_{m = -\infty}^\infty
  c_m \bigl(\pm g\bigr)^{m-1}\,, \quad &
  \del c(\pm1/g)=-
  \sum_{m = -\infty}^\infty
  (m-1) c_m \bigl(\pm g\bigr)^{m}\,.
\end{split}
\end{equation}
The leading term in the level expansion
of $\Psi^{(2)}_{(k,k')}$ in~(\ref{Psi2operator})
for $k=k'=0$ is given by
\begin{equation}
\begin{split}
& \int_0^1 dt(2+t)^{-L_0-1}
\Biggl[\frac{2g'}{(2+t)g^2}(\alpha_{-1}^+)^2c_1
-b_0(\alpha_{-1}^+)^2\frac{2}{g}c_0c_1
\Biggr]\ket{0} \\
& = 2
\int_0^1 dt\frac{\frac{g'}{2+t}-g}{(2+t)^2g^2}(\alpha_{-1}^+)^2c_1\ket{0}
=\frac{4}{3\sqrt{3}}(\alpha_{-1}^+)^2c_1\ket{0}\,.
\end{split}
\end{equation}
The above component of the
solution is exact to all orders in $\lambda$,
as it cannot receive contributions from
$\Psi^{(n)}$ with $n > 2$.
The coefficient was determined analytically
using
\emph{Mathematica}.

Let us now use the test state approach to determine this coefficient
for general $k$ and $k'$.
In other words, we are trying to determine $\beta_{k,k'}$ in
\be
\Psi_{k,k'}^{(2)}= \beta_{k,k'}e^{i(k_i+k_i')X^i(0)}(\alpha_{-1}^+)^2c_1\ket{0}+\dots
\ee
As always, the dots denote
higher-level contributions.
The appropriate test state $\phi_{k,k'}$
such that $\langle\phi_{k,k'},\Psi_{k,k'}^{(2)}\rangle
=\beta_{k,k'} \cdot (\hbox{vol})$
and its
BPZ conjugate
are given by
\begin{equation}\label{}
\begin{split}
|\phi_{k,k'} \rangle &=\frac{1}{2} e^{-i(k_i+k_i')X^i(0)}\, (\alpha_{-1}^-)^2 c_0 c_1 |0\rangle = \frac{1}{2}\,\Bigl(\frac{-1}{2\alpha'}\Bigr)\,
(\partial c) c \del X^-\del X^-
e^{-i(k_i+k_i')X^i}(0)\,  |0\rangle \,, \\
\langle \phi_{k,k'} |  &= \frac{1}{2} \lim_{y\to \infty}  \langle 0 | (\alpha_{1}^-)^2 c_{-1} c_0  e^{-i(k_i+k_i')X^i(y)}\,{\,1\over |y|^{2\alpha' (k+k')^2}} \,.
\end{split}
\end{equation}
The state $\phi_{k,k'}$ has
dimension
$\alpha'(k+k')^2+1$.
We can now evaluate
$\beta_{k,k'}$ as in the calculation of $\beta_{n+1}$ in~\S\ref{3.2}:
\begin{equation}\label{Psi2UHP}
\begin{split}
\beta_{k,k'} =
&\int_0^1 dt \, \frac{-1}{(2+t)2\alpha'}\Bigl(\frac{t f'(0) g'}{2+t}\Bigr)^{\alpha'(k^2+k'^2)}\Bigl\langle \frac{g'}{2+t} \Bigl\{\del X^+e^{ik_iX^i} (-g) \, h \circ \phi_{k,k'} (0) \, c \del X^+e^{ik'_iX^i} (g)\\
&\qquad+c\del X^+e^{ik_iX^i} (-g) \, h \circ \phi_{k,k'} (0) \,  \del X^+e^{ik'_iX^i} (g)\Bigr\}\\
 &\qquad+ c \del X^+e^{ik_iX^i} (-g) \, \Bigl[\hat{B} \, h\circ \phi_{k,k'}(0)\Bigr] \, c \del X^+e^{ik'_iX^i} (g) \, \Bigr\rangle_{\mathbb{H}}\\
=&~\frac{1}{2}
\Bigl(\frac{1}{2\alpha'}\Bigr)^2\int_0^1 dt \, \frac{h'(0)^{\alpha'(k+k')^2+1}}
{2+t}
\Bigl(\frac{t f'(0) g'}{2+t}\Bigr)^{\alpha'(k^2+k'^2)}\\
&\qquad\qquad\times
\Bigl\langle\del X^+e^{ik_iX^i} (-g) \,\,\del X^-\del X^-
e^{-i(k_i+k'_i) X^i}  (0) \,\,
\del X^+e^{ik'_iX^i} (g)\Bigr\rangle_m\\
&\qquad\qquad\times
\Bigl\langle
\frac{g'}{2+t} \,
\Bigl( (\del c) c(0) \, c(g)
+  c(-g) \, (\del c) c(0) \Bigr)
+ c(-g) \, c (0) \, c(g)
\Bigr\rangle_{g} \,,
\end{split}
\end{equation}
where we have again factored the correlator into
the matter and ghost sectors.
The matter contribution vanishes
unless each $\del X^+$ contracts with $\del X^-$,
and the ghost correlator has been calculated in~\S\ref{3.2}.
We therefore have
\begin{equation}
\begin{split}
\beta_{k,k'} =&\Bigl(\frac{1}{2\alpha'}\Bigr)^2\int_0^1 dt \, \frac{h'(0)^{\alpha'(k+k')^2+1}}
{2+t}
\Bigl(\frac{t f'(0) g'}{2+t}\Bigr)^{\alpha'(k^2+k'^2)}
2\Bigl(\frac{g'}{2+t}-g\Bigr)g^2\\
&\qquad\qquad\times
\Bigl\langle e^{ik_iX^i} (-g) \,\,
e^{-i(k_i+k'_i) X^i}  (0) \,\,
e^{ik'_iX^i} (g)\Bigr\rangle_m
\Bigl(\frac{(2\alpha')\eta^{+-}}{g^2}\Bigr)^2.
\end{split}
\end{equation}
We evaluate the remaining matter correlator
and use $h'(0) = \frac{1}{2+t}$
and $f'(0)=\frac{2}{\pi}$
to obtain
\begin{equation}\label{bkk'}
\begin{split}
\beta_{k,k'}
= 2
\int_0^1 dt \, (2+t)^{-\alpha'(k+k')^2-2}\Bigl(\frac{2t g'}{\pi(2+t)}\Bigr)^{\alpha'(k^2+k'^2)}
\Bigl(\frac{g'}{2+t}-g\Bigr)\frac{(2g)^{2\alpha'k\cdot k'}}{g^{2+2\alpha'(k+k')^2}}
\,.
\end{split}
\end{equation}
For general momenta the integral is
complicated,
but for $k = k' = 0$
we recover the  result from the operator expansion:
$\beta_{k=0,k'=0}=\frac{4}{3\sqrt{3}}\,.$
To summarize, our solution is
\begin{equation}
\begin{split}
\Psi=&~\lambda \int dk_i \, A(k_i) e^{ik_iX^i(0)}\alpha_{-1}^+c_1\ket{0} \\[0.5ex]
&+\lambda^2\Bigl(\int dk_idk'_i \, A(k_i)A(k'_i)\beta_{k,k'}e^{i(k_i+k_i')X^i(0)}(\alpha_{-1}^+)^2c_1\ket{0}+\dots\Bigr)
+ O(\lambda^3)
\end{split}
\end{equation}
with $\beta_{k,k'}$ given in~(\ref{bkk'}).

\sectiono{Solutions for marginal operators
with singular operator products}\label{section_four}

In the previous section,
we constructed analytic solutions for marginal deformations
when the operator $V$ has
regular operator products.
In this section
we generalize the construction to the case
where $V$ has
the following singular OPE with itself:
\begin{equation}
V(z) \, V(w) \sim \frac{1}{(z-w)^2} + \hbox{regular}.
\label{V-V}
\end{equation}

\subsection{Construction of $\Psi^{(2)}$}\label{4.1}

The string field $\Psi^{(2)}$ in (\ref{psi2_def}) is not well defined
when $V$ has
the OPE~(\ref{V-V}).
Let us  define a regularized string field
$\Psi^{(2)}_0$ as follows:
\begin{equation}
\langle \, \phi, \Psi^{(2)}_0 \, \rangle
= \int_{2 \epsilon}^1 dt \, \langle \, f \circ \phi (0) \,
c V (1) \, {\cal B} \, c V (1+t) \, \rangle_{{\cal W}_{1+t}} \,.
\end{equation}
The equation of motion
 is no longer satisfied by $\Psi^{(2)}_0$
because the surface term at $t=2 \epsilon$ in (\ref{taking})
is nonvanishing.
The BRST transformation of $\Psi^{(2)}_0$ is given by
\begin{equation}
\langle \, \phi, Q_B \Psi^{(2)}_0 \, \rangle
= {}- \langle \, \phi, \Psi^{(1)} \ast \Psi^{(1)} \, \rangle
+ \langle \, f \circ \phi (0) \,
c V (1) \, c V (1+2 \epsilon) \,
\rangle_{{\cal W}_{1+2 \epsilon}} \, ,
\end{equation}
and we see that  the second term on the right-hand side
violates the equation of motion.
Using the OPE
\begin{equation}
cV (-\epsilon) \, cV(\epsilon)
= \frac{1}{2 \epsilon} \, c \partial c (0) + O(\epsilon) \,,
\end{equation}
the  term violating the equation of motion can be written as
\begin{equation}
\langle \, f \circ \phi (0) \,
c V (1) \, c V (1+2 \epsilon) \,
\rangle_{{\cal W}_{1+2 \epsilon}}
= \frac{1}{2 \epsilon} \,
\langle \, f \circ \phi (0) \,
c \partial c (1+\epsilon) \,
\rangle_{{\cal W}_{1+2 \epsilon}} + O(\epsilon) \,.
\label{violation}
\end{equation}
Since the operator $c \partial c$
is the BRST transformation of $c$, we recognize that
the term~(\ref{violation}) is BRST exact
up to contributions which vanish as $\epsilon \to 0$.
This crucial property
makes it possible to satisfy the equation of motion
by adding a counterterm
to the regularized string field $\Psi^{(2)}_0$.
We define the counterterm
$\Psi^{(2)}_1$ by
\begin{equation}
\langle \, \phi, \Psi^{(2)}_1 \, \rangle
= {}- \frac{1}{2 \epsilon} \,
\langle \, f \circ \phi (0) \,
c (1+\epsilon) \, \rangle_{{\cal W}_{1+2 \epsilon}}\,.
\end{equation}
The sum of $\Psi^{(2)}_0$ and $\Psi^{(2)}_1$ then solves
the equation of motion in the limit $\epsilon \to 0$:
\begin{equation}
\lim_{\epsilon \to 0} \,
\langle \, \phi, Q_B \, (\Psi^{(2)}_0 + \Psi^{(2)}_1)
+ \Psi^{(1)} \ast \Psi^{(1)} \, \rangle = 0 \, .
\end{equation}
This is not yet the end of the story,
as we must also require that the solution
be finite as $\epsilon \to 0$.
Since $\Psi^{(1)} \ast \Psi^{(1)}$ is a finite state,
$Q_B \, (\Psi^{(2)}_0 + \Psi^{(2)}_1)$ is also finite
in the limit $\epsilon \to 0$. This implies that while
the state $\Psi^{(2)}_0 + \Psi^{(2)}_1$ can be divergent,
the divergent terms must be BRST closed.
It follows that
a finite solution can be obtained
by simply subtracting the divergent terms from
$\Psi^{(2)}_0 + \Psi^{(2)}_1$.
Let us isolate the divergent terms in $\Psi^{(2)}_0$.
Using the anticommutation relation $\{\, {\cal B}, c(z) \, \} = 1$,
the operator insertions in
$\Psi^{(2)}_0$
can be written as
\begin{equation}
\begin{split}
cV (1) \, {\cal B} \, cV (1+t)
& =  cV (1) \, V (1+t) - cV (1) \, cV (1+t) \, {\cal B} \\
& = \frac{1}{t^2} \, c(1)
- \frac{1}{t} \, c \partial c (1) \, {\cal B} + O(t^0) \,.
\end{split}
\end{equation}
Using the formula
\begin{equation}
\begin{split}
 \langle \, {\cal O}_1 (z_1) \, {\cal O}_2 (z_2) \, \ldots \,
{\cal O}_n (z_n) \,
\rangle_{{\cal W}_{\alpha + \delta \alpha}}
& = \langle \, {\cal O}_1 (z_1) \, {\cal O}_2 (z_2) \, \ldots \,
{\cal O}_n (z_n) \, \rangle_{{\cal W}_\alpha}\\
&~~+ \delta \alpha \, \langle \, {\cal O}_1 (z_1) \,
{\cal O}_2 (z_2) \, \ldots \, {\cal O}_n (z_n) \, {\cal L} \,
\rangle_{{\cal W}_\alpha} + O(\delta \alpha^2) \, ,
\end{split}
\end{equation}
valid for any set of operators ${\cal O}_i$, we find
\begin{equation}
\begin{split}
 \langle  f \circ \phi (0) \, cV (1) \, {\cal B} \, cV (1+t)
\rangle_{{\cal W}_{1+t}}
& = \frac{1}{t^2} \, \langle  f \circ \phi (0) \, c(1)
\rangle_{{\cal W}_{1}}
+ \frac{1}{t} \, \langle  f \circ \phi (0)
\bigl[ c(1)  {\cal L}
- c \partial c (1) {\cal B}  \bigr]
\rangle_{{\cal W}_{1}}  + O(t^0) \\[0.5ex]
&= \frac{1}{t^2} \, \langle \, f \circ \phi (0) \, c(1) \,
\rangle_{{\cal W}_{1}}
+ \frac{1}{t} \, \langle \, \phi,\, \psi_0' \, \rangle + O(t^0) \, ,
\end{split}
\end{equation}
where in the last equality we have
used
the expression for $\psi'_0$ \cite{Okawa:2006vm, ORZ}.
The first term on the right-hand side is not BRST closed.
After integration  over $t$, it gives
a divergent term of $O(1/\epsilon)$ which
is  precisely canceled
by the divergent term from $\Psi^{(2)}_1$, as expected.
The integral over $t$ of the second term gives
a divergent term of $O(\ln \epsilon)$
which  is not canceled but, as expected, is BRST closed.
(It is in fact BRST exact.)
If we define the counterterm $\Psi^{(2)}_2$ by
\begin{equation}
\Psi^{(2)}_2 = \ln (2 \epsilon) \, \psi'_0 \, ,
\end{equation}
we finally assemble a string field $\Psi^{(2)}$ that is
finite and satisfies the equation of motion as follows:
\begin{equation}
\Psi^{(2)} = \lim_{\epsilon \to 0} \,
\left[ \, \Psi^{(2)}_0 + \Psi^{(2)}_1 + \Psi^{(2)}_2 \, \right] \,.
\end{equation}
We can also write the solution as
\be \label{Psi2op}
\Psi^{(2)} = \lim_{\epsilon \to 0} \,
\left[ \,\Psi^{(2)}_0 -\frac{1}{\pi  \epsilon} \, c_1 |0 \rangle +  \ln (2 \epsilon) \, \psi'_0  +\frac{1}{\pi} \, L^+ \, c_1 |0\rangle \, \right] \,,
\ee
using the following operator expression for $\Psi^{(2)}_1$:
\begin{equation}
\Psi^{(2)}_1
= -\frac{1}{\pi  \epsilon} \, e^{-\epsilon L^+}  c_1 | 0 \rangle
= -\frac{1}{\pi  \epsilon} \, c_1 |0 \rangle
+\frac{1}{\pi} \, L^+ \, c_1 |0\rangle + O(\epsilon) \,.
\end{equation}

Our construction of $\Psi^{(2)}$ did not rely on any property of $V$ other
than the  OPE (\ref{V-V}).
The OPE (\ref{V-V}) is more restrictive than the generic OPE
of a dimension-one primary field.
For example, we may have
\begin{equation} \label{VVU}
V(z) \, V(w) \sim \frac{1}{(z-w)^2} + \frac{1}{z-w} \, U(w) \,,
\end{equation}
where $U(w)$ is some matter
primary field
of dimension one. In this
case, $V$ would not be exactly marginal. Indeed,
there must be a dimension-one primary field $\bar U$
such that $\langle \bar U(z) U(0) \rangle = 1/z^2$.
The OPE (\ref{VVU}) then implies that
the three-point function $\langle V\, V\, \bar U \rangle$ is
nonvanishing,
while a necessary condition for the exact marginality of $V$
is the vanishing of $\langle V V W \rangle $  for all
dimension-one primary fields
$W$. (See, for example, \cite{Recknagel:1998ih}.)
Thus we expect  that our construction of $\Psi^{(2)}$ should
not go through if the OPE takes the form (\ref{VVU}). Let us see this explicitly.
In this case (\ref{violation}) is replaced by
\begin{equation}
\begin{split}
\langle \, f \circ \phi (0) \,
c V (1) \, c V (1+2 \epsilon) \,
\rangle_{{\cal W}_{1+2 \epsilon}}
& = \frac{1}{2 \epsilon} \,
\langle \, f \circ \phi (0) \,
c \partial c (1+\epsilon) \,
\rangle_{{\cal W}_{1+2 \epsilon}} \\
& \qquad + \langle \, f \circ \phi (0) \,
c \partial c U (1+\epsilon) \,
\rangle_{{\cal W}_{1+2 \epsilon}} + O(\epsilon) \,.
\end{split}
\end{equation}
The second term on the right-hand side is finite
in the limit $\epsilon \to 0 \,$.
The operator $c \partial c U$ is BRST closed,
but it is {\it not} BRST exact.
Therefore the equation of motion cannot be satisfied
by adding a counterterm.

\subsection{Gauge condition, $L$ eigenstates, and divergence structure}

All the terms of $\Psi^{(2)}$ in (\ref{Psi2op})
are annihilated by $B$ except $L^+ c_1 |0 \rangle$:
\be
B L^+ c_1 | 0 \rangle = [B, L^+] c_1 |0 \rangle = B^+ c_1 |0 \rangle \neq 0 \,.
\ee
Thus, rather curiously, $\Psi^{(2)}$  violates the Schnabl gauge condition.
It appears that this violation is intrinsic.
While we can add an arbitrary BRST closed state $Z$ to $\Psi^{(2)}$,
we believe that no choice of $Z$ can restore
the Schnabl gauge condition.
Indeed, assume that such
a state $Z$ exists:
\be
B (  L^+ c_1 | 0\rangle + Z) = 0 \, , \quad Q_B Z = 0 \, .
\ee
Acting with $Q_B$ on this equation, we find that $Z$ must satisfy
\be \label{LZ}
L \, Z = - Q_B B L^+ c_1 | 0\rangle \, .
\ee
Note that
while the left-hand side
is in the image of $L$,
the right-hand side is in the kernel
of $L$
because $[L, Q_B] = [L, B]=0$ and $L L^+ c_1 | 0 \rangle = 0 \,$.
We believe that
(\ref{LZ}) has no solution for $Z$,
though we do not have a proof.\footnote{If an operator is diagonalizable,
 its kernel and  its image have no
nontrivial
overlap.  Since $L$ is non-hermitian,
it is not a priori clear
if
it can be diagonalized.
In principle a state $Z$
solving (\ref{LZ}) may exist if $L$ has
a suitable Jordan structure, but we find this unlikely.}

This obstruction in preserving
the Schnabl gauge condition when $V$ has
the singular OPE~(\ref{V-V})
is rather unexpected.
To gain some insight, let us reconsider
the situation in Siegel gauge.   In Siegel gauge the equations of motion (\ref{PertEom})
are solved by setting
\be
\Psi^{(n)} = \frac{b_0}{L_0}  \Phi^{(n)} \,.
\ee
It turns out that the right-hand side is well defined
and thus manifestly obeys the gauge condition
because $\Phi^{(n)}$ has no overlap with states in the kernel of $L_0$.
When the equations of motion have a solution,
$\Phi^{(n)}$ is
a BRST-exact state
of ghost number two.
The only BRST-exact state
of ghost number two
in the kernel of $L_0$ is $ Q_B c_0 |0 \rangle = 2 c_1 c_{-1} | 0 \rangle$.
We are claiming that  $\Phi^{(n)}$ has no overlap with $c_1 c_{-1}  |0 \rangle$.
This is shown using twist symmetry in the ghost sector.
For a generic state in the Fock space
\be
|\phi \rangle = \{ {\rm matter \; oscillators} \} \; b_{-m_j} \cdots  b_{-m_1} c_{-n_k} \cdots c_{-n_1} | 0 \rangle \,, \qquad
m_i \geq 2 \,, \; n_i \geq -1 \, ,
\ee
the ghost-twist eigenvalue
is defined
to be
\be
1+ \sum_{i=1}^j m_i + \sum_{i=1}^k n_i \, \quad  ({\rm mod} \; 2) \,.
\ee
The linearized solution $\Psi^{(1)}$ is even under ghost twist,
which implies that  $\Phi^{(2)} = - \Psi^{(1)} * \Psi^{(1)}$ is also even.
On the other hand, the problematic state $c_1 c_{-1} | 0 \rangle$ is odd. This shows that $\Phi^{(2)}$ has no overlap
with it.  A little inductive argument can be used to extend
this result to
$\Phi^{(n)}$ with $n > 2$.
Assuming that
all the states $\Psi^{(k)}$ with $k<n$
are even, we see that $\Phi^{(n)}$, which consists of
symmetrized star products
of the states $\Psi^{(k)}$ with $k < n$,
is also even. Hence there is no obstruction in finding $\Psi^{(n)} = \frac{b_0}{L_0} \Phi^{(n)}$.
The operator $b_0/L_0$ preserves twist, so $\Psi^{(n)}$ is even,
and the induction can proceed
to the next step.

We now perform a similar analysis for the case of Schnabl gauge.
The formal solution
\be
\Psi^{(n)} = \frac{B}{L}  \Phi^{(n)} \,
\ee
is well defined if and only if
$\Phi^{(n)}$ has no overlap
with states in the kernel of $L$.
 While we do not have a complete
understanding of the spectrum of $L$, we will find a consistent
picture by assuming that $\Phi^{(n)}$ can be expanded
in a sum of
eigenstates of $L$ with integer eigenvalues $L \geq -1$.\footnote
{Here and in what follows we use $L$ to denote the eigenvalue
of $L$ as well.}
We can systematically enumerate the $L$ eigenstates
that have ghost number two and are BRST exact
within a subspace of states
which can appear in the expansion of $\Phi^{(n)}$.
It will be sufficient to focus on states
with $L\leq 0$. We believe that the only such states
are as follows.
\begin{itemize}

\item $L=-1$: the state   $c_1 c_0 | 0 \rangle = Q_B  c_1 |0 \rangle $.

\item $L=0$: the state $c_{1} c_{-1} |0 \rangle = \frac{1}{2} Q_B c_0 | 0 \rangle$.

\item $L=0$: the state $ L^+ c_1 c_0 | 0 \rangle = Q_B  L^+ c_1 |0 \rangle  $.

\end{itemize}
Contrasting the kernel of $L$ with the kernel of $L_0$, we see the surprising appearance of the extra state
$L^+ c_1 c_0 |0 \rangle$.
Since this state is {\it even} under ghost twist,
it can a priori appear in $\Phi^{(n)}$.
The first state $c_{1} c_{-1} |0 \rangle$ with $L=0$
cannot appear,
as we have argued before.
We can write the following
ansatz for a finite $\Phi^{(n)}$:
\be
\Phi^{(n)} = \alpha^{(n)}  c_1 c_0 |0 \rangle + \beta^{(n)} L^+ c_1 c_0 |0 \rangle + \Phi^{(n)}_> \, ,
\ee
where $\Phi^{(n)}_>$
only contains terms with positive eigenvalues of $L$.
The most general  $\Psi^{(n)}$ that
satisfies the equation $Q_B \Psi^{(n)} = \Phi^{(n)}$  is the manifestly finite string field
\be
\label{findsdfh}
\Psi^{(n)} = \alpha^{(n)}  c_1 |0 \rangle + \beta^{(n)} L^+ c_1 |0 \rangle  + \frac{B}{L} \Phi^{(n)}_>
+ ({\rm BRST~closed})\,.
\ee
If $\beta^{(n)} \neq 0$, the term $L^+ c_1 |0 \rangle$ violates the gauge condition.
In the following we will not write
the BRST-closed term
that plays no role.

We are now going to establish a precise relationship between the violation
of the gauge condition and the divergences
that can arise
in the Schwinger representation of
the action of $B/L$
when
the matter operator
has singular operator products.
When $B/L$ acts on $\Phi^{(n)}_>$,
we can use its Schwinger representation
\be
\frac{B}{L} = \lim_{\Lambda \to \infty} \int_0^\Lambda dt  \, B e^{-t L}  = \frac{B}{L} - \lim_{\Lambda \to \infty} e^{- \Lambda L}\frac{B }{L}  \, ,
\ee
since the boundary term vanishes in the limit.
Thus we rewrite (\ref{findsdfh}) as
\begin{eqnarray}
\Psi^{(n)} &  = & \alpha^{(n)}  c_1 |0 \rangle + \beta^{(n)} L^+ c_1 |0\rangle +  \lim_{\Lambda \to \infty} \int_0^\Lambda dt\,
 B e^{-t L}  (  \Phi^{(n)} -  \alpha^{(n)}  c_1 c_0 |0 \rangle - \beta^{(n)} L^+ c_1 c_0 |0  \rangle  ) \nonumber \\
 & = &   \lim_{\Lambda \to \infty}  \left[  \left(  \int_0^\Lambda dt\,
  B e^{-t L}    \Phi^{(n)}\right)  +  e^{\Lambda}   \alpha^{(n)}  c_1 |0 \rangle - \Lambda \beta^{(n)}  B L^+ c_1 c_0 |0\rangle  \right]
 +  \beta^{(n)} L^+ c_1 |0\rangle \,.\label{genstructure}
 \end{eqnarray}
Note that we have
 \be
 B  L^+ c_1 c_0 |0\rangle
 = \pi  \psi'_0 \,.
 \ee
Since the string field  $\Psi^{(n)}$ is  finite,
 we see that
 \be \label{integral}
\int_0^\Lambda dt\,
  B e^{-t L}    \Phi^{(n)} = - e^{\Lambda}   \alpha^{(n)}  c_1 |0 \rangle +  \Lambda \, \pi \beta^{(n)}  \,  \psi_0'  + {\rm finite} \,.
\ee
We have thus learned that the divergences of the integral on the left-hand side,
which performs the naive inversion of $Q_B$ on $\Phi^{(n)}$, are directly related to the $L=-1$ and $L=0$ eigenstates in the decomposition of $\Phi^{(n)}$. Moreover, the coefficient of
the divergence of $O(\Lambda)$
is correlated with the coefficient of
the Schnabl-gauge
violating term $ L^+ c_1 |0\rangle$.

 The divergences in  (\ref{integral})  can only arise from the collision of the $cV$
 insertions on the boundary of the world-sheet. If $V$ has
regular operator products,
all integrals are manifestly finite,  $\alpha^{(n)} = \beta^{(n)} = 0$ for any $n$, $\Psi^{(n)}$ satisfies
the Schnabl gauge condition, and the naive prescription $Q_B^{-1} = B/L$
is adequate to handle this case, as discussed in section 3.
On the other hand, if $V$ has a
singular OPE with itself,
(\ref{genstructure}) severely constrains the structure of the result.
Let us look  at the case of $\Psi^{(2)}$.
To begin with,
note that the integral
\be \label{Psi2inte}
\int_0^\Lambda  dt\,
  B e^{-t L} \Phi^{(2)}
\ee
is in fact the regularized $\Psi_0^{(2)}$
with the identification
$\Lambda =-\ln( 2 \epsilon)$.
Substituting this in (\ref{genstructure}),
our general analysis predicts
\be
\Psi^{(2)} = \lim_{\epsilon \to 0} \,
\left[ \,\Psi^{(2)}_0  +\frac{ \alpha^{(2) }}{ 2 \epsilon} \, c_1 |0 \rangle +  \ln (2 \epsilon) \pi \, \beta^{(2)}\, \psi'_0  +\beta^{(2)} \, L^+ \, c_1 |0\rangle \, \right]
\ee
in complete agreement
with the explicit result (\ref{Psi2op})
with
$\alpha^{(2)} =-2/\pi $ and $\beta^{(2)} = 1/\pi$.

The analysis can be extended
to $\Psi^{(n)}$ with $n > 2$.
An interesting
simplification occurs if $V =  i \sqrt{\frac{2}{\alpha'}} \partial X$.
Since the number of $\partial X$ is conserved mod $2$
under Wick contractions,
the coefficients $\alpha^{(n)}$ and $\beta^{(n)}$ are zero
for odd $n$.
It follows that for odd $n$ the integral (\ref{integral}) is finite.
In particular, we expect that
for $V =   i \sqrt{\frac{2}{\alpha'}} \partial X $
the most general $\Psi^{(3)}$ is given by
\be \label{Psi3gene}
\Psi^{(3)} = -\lim_{\Lambda \to \infty} \int_0^{\Lambda}  dt  \, Be^{- t  L}  \left(   \Psi^{(1)} * \Psi^{(2)} + \Psi^{(2)} * \Psi^{(1)} \right)  +
({\rm BRST~closed}) \, ,
\ee
where the $\Lambda \to \infty$ limit is guaranteed to be finite.

While $\Psi^{(3)}$ may be obtained
this way (setting
the arbitrary BRST closed terms to zero
and performing the integral by brute force),
in the following subsection
we will follow a route analogous to the one
in \S\ref{4.1}.
We will start with a regularized $\Psi^{(3)}_0$ and systematically look for
counterterms
such that the final state
$\Psi^{(3)}$ satisfies the equation of motion and is finite.
The arguments in this section
strongly suggest
that a finite string
field $\Psi^{(n)}$ satisfying the equation of motion
exists for all $n$ and it can be written as a regularized string
field plus counterterms.

\subsection{Construction of $\Psi^{(3)}$}

In this subsection we perform an explicit  construction of $\Psi^{(3)}$ for $V$ with the OPE (\ref{V-V}).
The first step is to regularize (\ref{psingeometric})
and define $\Psi^{(3)}_0$ by
\begin{equation}
\langle \phi, \Psi^{(3)}_0\rangle
= \int_{2 \epsilon}^1 dt_1 \int_{2 \epsilon}^1 dt_2 \,
\langle \, f \circ \phi (0) \,
c V (1) \, {\cal B} \, c V (1+t_1) \, {\cal B} \, c V (1+t_1+t_2) \,
\rangle_{{\cal W}_{1+t_1+t_2}} \,.
\label{Psi^(3)_0}
\end{equation}
The BRST transformation of $\Psi^{(3)}_0$ is given by
\begin{equation}
\langle \phi, Q_B\Psi^{(3)}_0\rangle
= {}- \langle \, \phi, \Psi^{(1)} \ast \Psi^{(2)}_0
+ \Psi^{(2)}_0 \ast \Psi^{(1)} \, \rangle
+ R_1 + R_2 \,, \\
\end{equation}
where
\begin{equation}
\begin{split}
R_1 & = \int_{2 \epsilon}^1 dt_2 \, \langle \, f \circ \phi (0) \,
c V (1) \, c V (1+2 \epsilon) \, {\cal B} \,
c V (1+2 \epsilon+t_2) \,
\rangle_{{\cal W}_{1+t_2+2 \epsilon}} \,, \\[0.5ex]
R_2 & = \int_{2 \epsilon}^1 dt_1 \,
\langle \, f \circ \phi (0) \,
c V (1) \, {\cal B} \, c V (1+t_1) \, c V (1+t_1+2 \epsilon) \,
\rangle_{{\cal W}_{1+t_1+2 \epsilon}} \,.
\end{split}
\end{equation}
As in the case of $Q_B \Psi^{(2)}_0$,
the contributions $R_1$ and $R_2$ from the surface terms
at $t_1 = 2 \epsilon$ and at $t_2 = 2 \epsilon$, respectively,
are nonvanishing.
We also need to reproduce
${}- \Psi^{(1)} \ast \Psi^{(2)}_1 - \Psi^{(2)}_1 \ast \Psi^{(1)}$
and
${}- \Psi^{(1)} \ast \Psi^{(2)}_2 - \Psi^{(2)}_2 \ast \Psi^{(1)}$
to satisfy the equation of motion.
It is not difficult to realize that the BRST transformation
of $\Psi^{(3)}_1$ defined by
\begin{equation}
\Psi^{(3)}_1 = {}- \int_{2 \epsilon}^1 dt_1 \,
\Psi^{(1)} \ast B^+_L \, e^{(1-t_1) L^+_L} \, \Psi^{(2)}_1
- \int_{2 \epsilon}^1 dt_2 \,
\Psi^{(2)}_1 \ast B^+_L \, e^{(1-t_2) L^+_L} \,
\Psi^{(1)}
\end{equation}
cancels the divergent terms from the OPE's
of $c V (1) \, c V (1+2 \epsilon)$ in $R_1$
and of $c V (1+t_1) \, c V (1+t_1+2 \epsilon)$ in $R_2$
and reproduces
${}- \Psi^{(1)} \ast \Psi^{(2)}_1 - \Psi^{(2)}_1 \ast \Psi^{(1)}$.
We also introduce $\Psi^{(3)}_2$ defined by
\begin{equation}
\Psi^{(3)}_2 = {}- \int_{2 \epsilon}^1 dt_1 \,
\Psi^{(1)} \ast B^+_L \, e^{(1-t_1) L^+_L} \, \Psi^{(2)}_2
- \int_{2 \epsilon}^1 dt_2 \,
\Psi^{(2)}_2 \ast B^+_L \, e^{(1-t_2) L^+_L} \,
\Psi^{(1)}
\end{equation}
so that its BRST transformation reproduces
${}- \Psi^{(1)} \ast \Psi^{(2)}_2 - \Psi^{(2)}_2 \ast \Psi^{(1)}$.

However, this is not the whole story.
First, when $t_2$ in $R_1$ is of $O(\epsilon)$,
three $V$'s are simultaneously close
so that we cannot simply replace two of them
by the most singular term of the OPE.
The same remark applies to $R_2$ when $t_1$ is of $O(\epsilon)$.
Secondly, while the contributions from the surface terms
at $t_1 = 2 \epsilon$ or at $t_2 = 2 \epsilon$
in the calculation of $Q_B \Psi^{(3)}_2$
turn out to vanish in the limit $\epsilon \to 0$,
the corresponding contributions
in the calculation of $Q_B \Psi^{(3)}_1$
turn out to be {\it finite} and {\it not} BRST exact.
These contributions have to be canceled
in order for the equation of motion to be satisfied.

We thus need to calculate
$R_1$, $R_2$, $Q_B \Psi^{(3)}_1$, and $Q_B \Psi^{(3)}_2$.
The calculations of $Q_B \Psi^{(3)}_1$ and $Q_B \Psi^{(3)}_2$
are universal for any $V$ which has the OPE (\ref{V-V}),
while those of $R_1$ and $R_2$ are not.
Let us begin with $Q_B \Psi^{(3)}_1$.
It is convenient to use the CFT description of $\Psi^{(3)}_1$
given by
\begin{equation}
\begin{split}
\langle \phi, \Psi^{(3)}_1\rangle =& {}- \frac{1}{2 \epsilon} \, \int_{2 \epsilon}^1 dt_1 \,
\langle \, f \circ \phi (0) \,
c V (1) \, {\cal B} \, c (1+t_1+\epsilon) \,
\rangle_{{\cal W}_{1+t_1+2 \epsilon}} \\[1.0ex]
& {}- \frac{1}{2 \epsilon} \, \int_{2 \epsilon}^1 dt_2 \,
\langle \, f \circ \phi (0) \,
c (1+\epsilon) \, {\cal B} \, c V (1+t_2+2 \epsilon) \,
\rangle_{{\cal W}_{1+t_2+2 \epsilon}} \,.
\end{split}
\end{equation}
The BRST transformation of $\Psi^{(3)}_1$ is
\begin{equation}
\langle \phi, Q_B \Psi^{(3)}_1\rangle
= {}- \langle \, \phi,  \Psi^{(1)} \ast \Psi^{(2)}_1
+ \Psi^{(2)}_1 \ast \Psi^{(1)} \, \rangle
+ \widetilde{R}_1 + \widetilde{R}_2 + \widetilde{R}_3 \,, \\
\end{equation}
where
\begin{equation}
\begin{split}
\widetilde{R}_1 & = - \frac{1}{2 \epsilon} \,
\int_{2 \epsilon}^1 dt_2 \, \langle \, f \circ \phi (0) \,
c \partial c (1+\epsilon) \, {\cal B} \, c V (1+t_2+2 \epsilon) \,
\rangle_{{\cal W}_{1+t_2+2 \epsilon}} \,, \\
\widetilde{R}_2 & = - \frac{1}{2 \epsilon} \,
\int_{2 \epsilon}^1 dt_1 \, \langle \, f \circ \phi (0) \,
c V (1) \, {\cal B} \, c \partial c (1+t_1+\epsilon) \,
\rangle_{{\cal W}_{1+t_1+2 \epsilon}} \,, \\
\widetilde{R}_3 & = - \frac{1}{2 \epsilon} \,
\langle \, f \circ \phi (0) \, c V (1) \, c (1+3 \epsilon) \,
\rangle_{{\cal W}_{1+4 \epsilon}}
- \frac{1}{2 \epsilon} \, \langle \, f \circ \phi (0) \,
c (1+\epsilon) \, c V (1+4 \epsilon) \,
\rangle_{{\cal W}_{1+4 \epsilon}} \,.
\end{split}
\end{equation}
As we mentioned earlier,
the BRST transformation of $\Psi^{(3)}_1$ reproduces
$- \Psi^{(1)} \ast \Psi^{(2)}_1 - \Psi^{(2)}_1 \ast \Psi^{(1)}$,
and $\widetilde{R}_1$ and $\widetilde{R}_2$ cancel
part of $R_1$ and $R_2$, respectively.
The last term $\widetilde{R}_3$ is finite
in the limit $\epsilon \to 0$ and not BRST exact:
\begin{equation}
\widetilde{R}_3 = {}-3 \, \langle \, f \circ \phi (0) \,
c \partial c V (1) \, \rangle_{{\cal W}_{1}}
+ O(\epsilon) \,.
\end{equation}

Let us next calculate
$Q_B \Psi^{(3)}_2$.
It is again convenient to use
the CFT description of $\Psi^{(3)}_2$:
\begin{equation}
\begin{split}
\langle \, \phi, \Psi^{(3)}_2 \, \rangle
= &\quad \ln (2 \epsilon)  \int_{2 \epsilon}^1 dt_1 \,
\langle \, f \circ \phi (0) \,
c V (1) \, {\cal B} \,\,
Q_B \cdot [ \, {\cal B} \, c (1+t_1) \, ] \,
\rangle_{{\cal W}_{1+t_1}} \\[1.0ex]
&+ \ln (2 \epsilon) \int_{2 \epsilon}^1 dt_2 \,
\langle \, f \circ \phi (0) \,\,
Q_B \cdot [ \, {\cal B} \, c (1) \, ] \,\,
{\cal B} \, c V (1+t_2) \, \rangle_{{\cal W}_{1+t_2}} \,.
\end{split}
\end{equation}
The BRST transformation of  $\Psi^{(3)}_2$ is given by
\begin{equation}
\begin{split}
\langle \, \phi, Q_B \Psi^{(3)}_2 \, \rangle
= & {}- \, \langle \, \phi, \,
\Psi^{(1)} \ast \Psi^{(2)}_2
+ \Psi^{(2)}_2 \ast \Psi^{(1)} \, \rangle \\[1.0ex]
&{}- \ln (2 \epsilon) \, \langle \, f \circ \phi (0) \,\,
Q_B \cdot [ \, c V (1) \, {\cal B} \, c (1+2 \epsilon) \, ] \,
\rangle_{{\cal W}_{1+2 \epsilon}} \\[1.0ex]
&+ \ln (2 \epsilon) \, \langle \, f \circ \phi (0) \,\,
Q_B \cdot [ \, {\cal B} \, c (1) \,
c V (1+2 \epsilon) \, ] \, \rangle_{{\cal W}_{1+2 \epsilon}} \,.
\end{split}
\end{equation}
Since the BRST transformations of
$c V (1) \, {\cal B} \, c (1+2 \epsilon)$
and ${\cal B} \, c (1) \, c V (1+2 \epsilon)$
are both of $O(\epsilon)$,
the last two terms vanish in the limit $\epsilon \to 0$.
We have thus shown that
\begin{equation}
\lim_{\epsilon \to 0} \,
\langle \, \phi, Q_B \Psi^{(3)}_2
+ \Psi^{(1)} \ast \Psi^{(2)}_2
+ \Psi^{(2)}_2 \ast\Psi^{(1)} \, \rangle = 0 \,.
\end{equation}

To summarize, we have seen that the BRST transformation of
$\Psi^{(3)}_0 + \Psi^{(3)}_1 + \Psi^{(3)}_2$
reproduces
$- \Psi^{(1)} \ast \Psi^{(2)} - \Psi^{(2)} \ast \Psi^{(1)}$
with
$\Psi^{(2)} = \Psi^{(2)}_0 + \Psi^{(2)}_1 + \Psi^{(2)}_2$,
and there are remaining terms $R_1$, $R_2$,
$\widetilde{R}_1$, $\widetilde{R}_2$, and $\widetilde{R}_3$.
We now calculate $R_1$ and $R_2$.
These terms involve a triple operator product of $V$'s
and the results depend on $V$.
We choose
\begin{equation}
V(z) = i \sqrt{\frac{2}{\alpha'}} \, \partial X (z) \,,
\end{equation}
which is {\it exactly} marginal.
With this choice of $V$, the triple operator product of $V$'s
on ${\cal W}_{n-1}$ is
\begin{equation}
\begin{split}
V(z_1) V(z_2) V(z_3)
&= G_{n-1} (z_1-z_2) \, V(z_3) + G_{n-1} (z_1-z_3) \, V(z_2)
+ G_{n-1} (z_2-z_3) \, V(z_1) \\[1.0ex]
& \qquad + :\hskip-2pt  V(z_1) V(z_2) V(z_3) \hskip-2pt : \,,
\end{split}
\label{V-V-V}
\end{equation}
where $G_{n-1}$ is the propagator on ${\cal W}_{n-1}$:
\begin{equation}
G_{n-1} (z) = \frac{\pi^2}{n^2} \,
\left[ \, \sin \frac{\pi z}{n} \, \right]^{-2}
= \frac{1}{z^2} + O(z^0) \,.
\end{equation}
The normal-ordered term in (\ref{V-V-V}) does not
contribute in the calculations of $R_1$ and $R_2$
in the limit $\epsilon \to 0 \,$.
The term with $V(1)$ and $V(1+2 \epsilon)$ contracted in $R_1$
cancels $\widetilde{R}_1$:
\begin{equation}
\lim_{\epsilon \to 0} \, \biggl[ \,
\int_{2 \epsilon}^1 dt_2 \,
G_{1+t_2+2 \epsilon} (2 \epsilon) \,
\langle \, f \circ \phi (0) \,
c (1) \, c (1+2 \epsilon) \, {\cal B} \,
c V (1+2 \epsilon+t_2) \,
\rangle_{{\cal W}_{1+t_2+2 \epsilon}}
+ \widetilde{R}_1 \, \biggr] = 0 \,.
\end{equation}
The remaining two terms are finite in the limit $\epsilon \to 0$:
\begin{equation}
\begin{split}
& \lim_{\epsilon \to 0} \, \biggl[ \,
\int_{2 \epsilon}^1 dt_2 \,
G_{1+t_2+2 \epsilon} (t_2) \,
\langle \, f \circ \phi (0) \,
c V (1) \, c (1+2 \epsilon) \, {\cal B} \,
c (1+2 \epsilon+t_2) \,
\rangle_{{\cal W}_{1+t_2+2 \epsilon}} \\
& \qquad \quad {}+ \int_{2 \epsilon}^1 dt_2 \,
G_{1+t_2+2 \epsilon} (t_2 + 2 \epsilon) \,
\langle \, f \circ \phi (0) \,
c (1) \, c V (1+2 \epsilon) \, {\cal B} \,
c (1+2 \epsilon+t_2) \,
\rangle_{{\cal W}_{1+t_2+2 \epsilon}} \, \biggr] \\
& = \frac{3}{2} \, \langle \, f \circ \phi (0) \,
c \partial c V(1) \, \rangle_{{\cal W}_{1}} .
\end{split}
\end{equation}
We therefore have
\begin{equation}
\lim_{\epsilon \to 0} \, \left[ \, R_1 + \widetilde{R}_1 \, \right]
= \frac{3}{2} \, \langle \, f \circ \phi (0) \,
c \partial c V(1) \, \rangle_{{\cal W}_{1}} .
\end{equation}
The calculation of $R_2$ is parallel, and we obtain
\begin{equation}
\lim_{\epsilon \to 0} \, \left[ \, R_2 + \widetilde{R}_2 \, \right]
= \frac{3}{2} \, \langle \, f \circ \phi (0) \,
c \partial c V(1) \, \rangle_{{\cal W}_{1}} .
\end{equation}
The sum of the five remaining terms vanishes
in the limit $\epsilon \to 0$:
\begin{equation}
\lim_{\epsilon \to 0} \, \left[ \,
R_1 + R_2 + \widetilde{R}_1 + \widetilde{R}_2 + \widetilde{R}_3 \,
\right] = 0 \,.
\end{equation}
We have thus shown
\begin{equation}
\lim_{\epsilon \to 0} \,
\langle \, \phi, Q_B \, [ \, \Psi^{(3)}_0 + \Psi^{(3)}_1 \, ]
+ \Psi^{(1)} \ast [ \, \Psi^{(2)}_0 + \Psi^{(2)}_1 \, ]
+ [ \, \Psi^{(2)}_0 + \Psi^{(2)}_1 \, ] \ast\Psi^{(1)} \,
\rangle = 0
\end{equation}
and
\begin{equation}
\lim_{\epsilon \to 0} \,
\langle \, \phi,
Q_B \, [ \, \Psi^{(3)}_0 + \Psi^{(3)}_1 + \Psi^{(3)}_2 \, ]
+ \Psi^{(1)} \ast \Psi^{(2)} + \Psi^{(2)} \ast\Psi^{(1)} \,
\rangle = 0 \,.
\end{equation}
The sum of the five terms did not have to vanish
in the limit $\epsilon \to 0 \,$, but it had to be BRST exact
to satisfy the equation of motion
by adding a counterterm.
In particular, the coefficient
in front of
$\langle \, f \circ \phi (0) \,
c \partial c V(1) \, \rangle_{{\cal W}_{1}}$ had to vanish.
We found that $\widetilde{R}_3$ from $\Psi^{(3)}_1$
is nontrivially canceled by contributions from $\Psi^{(3)}_0$.

Let us next study the divergent terms of $\Psi^{(3)}_0$.
The triple operator product of $V$'s in (\ref{Psi^(3)_0})
can be written as follows:
\begin{equation}
\begin{split}
& V(1) \, V(1+t_1) \, V(1+t_1+t_2) \\
& = G_{1+t_1+t_2} (t_2) \, V(1)
+ G_{1+t_1+t_2} (t_1) \, V(1+t_1+t_2) \\
& \qquad {}+ G_{1+t_1+t_2} (t_1 + t_2) \, V(1+t_1) \,
{}+ : V(1) \, V(1+t_1) \, V(1+t_1+t_2) : \,.
\end{split}
\end{equation}
Note that no further divergence appears
when remaining operators collide.
The contribution from the normal-ordered product
in the last line is obviously finite.
The divergent terms from the first two terms
on the right-hand side are canceled
by the divergent terms from $\Psi^{(3)}_1$ and $\Psi^{(3)}_2$.
The contribution from the third term
on the right-hand side is
\begin{equation}
\begin{split}
& \int_{2 \epsilon}^1 dt_1 \int_{2 \epsilon}^1 dt_2 \,
\left( \, \frac{\pi}{t_1+t_2+2} \, \right)^2 \left[ \,
\sin \frac{\pi \, (t_1+t_2)}{t_1+t_2+2} \, \right]^{-2} \\
& \qquad \qquad \qquad \quad {}\times
\langle \, f \circ \phi (0) \,
c(1) \, {\cal B} \, c V (1+t_1) \, {\cal B} \, c(1+t_1+t_2) \,
\rangle_{{\cal W}_{1+t_1+t_2}} .
\end{split}
\end{equation}
This contains a divergent term
$-\ln ( 4 \epsilon ) \, \langle \, f \circ \phi (0) \,
cV (1) \, \rangle_{{\cal W}_1}$,
which comes from the most singular term $1/(t_1+t_2)^2$
in the region where $t_1$ and $t_2$
are simultaneously of $O(\epsilon)$.
Note that the divergent term is proportional to $\Psi^{(1)}$
and thus BRST closed, as expected.
Therefore, if we define
\begin{equation}
\Psi^{(3)} = \lim_{\epsilon \to 0} \,
\bigl[ \, \Psi^{(3)}_0 + \Psi^{(3)}_1
+ \Psi^{(3)}_2 + \Psi^{(3)}_3 \, \bigr] \,,
\end{equation}
where
\begin{equation}
\Psi^{(3)}_3 = \ln ( 4 \epsilon ) \, \Psi^{(1)} \,,
\end{equation}
$\Psi^{(3)}$ is finite and satisfies the equation of motion:
\begin{equation}
\langle \, \phi, \, Q_B \Psi^{(3)}
+ \Psi^{(1)} \ast \Psi^{(2)} + \Psi^{(2)} \ast \Psi^{(1)} \,
\rangle = 0 \,.
\end{equation}
An explicit form of $\Psi^{(3)}$ is given by
\begin{equation}
\begin{split}
\langle \, \phi, \Psi^{(3)} \, \rangle
= \lim_{\epsilon \to 0} \, \biggl[ \,
& \int_{2 \epsilon}^1 dt_1 \int_{2 \epsilon}^1 dt_2 \,
\langle \, f \circ \phi (0) \,
c V (1) \, {\cal B} \, c V (1+t_1) \, {\cal B} \, c V (1+t_1+t_2) \,
\rangle_{{\cal W}_{1+t_1+t_2}} \\[1.5ex]
& {}- \frac{1}{2 \epsilon} \, \int_{2 \epsilon}^1 dt_1 \,
\langle \, f \circ \phi (0) \,
c V (1) \, {\cal B} \, c (1+t_1+\epsilon) \,
\rangle_{{\cal W}_{1+t_1+2 \epsilon}} \\
& {}- \frac{1}{2 \epsilon} \, \int_{2 \epsilon}^1 dt_2 \,
\langle \, f \circ \phi (0) \,
c (1+\epsilon) \, {\cal B} \, c V (1+t_2+2 \epsilon) \,
\rangle_{{\cal W}_{1+t_2+2 \epsilon}} \\[1.5ex]
& {}+ \ln (2 \epsilon) \int_{2 \epsilon}^1 dt_1 \,
\langle \, f \circ \phi (0) \,
c V (1) \, {\cal B} \,\,
Q_B \cdot [ \, {\cal B} \, c (1+t_1) \, ] \,
\rangle_{{\cal W}_{1+t_1}} \\
& {}+ \ln (2 \epsilon) \int_{2 \epsilon}^1 dt_2 \,
\langle \, f \circ \phi (0) \,\,
Q_B \cdot [ \, {\cal B} \, c (1) \, ] \,\,
{\cal B} \, c V (1+t_2) \, \rangle_{{\cal W}_{1+t_2}} \\[1.5ex]
& {}+ \ln (4 \epsilon) \, \langle \, f \circ \phi (0) \,
c V (1) \, \rangle_{{\cal W}_1} \, \biggr] \,.
\end{split}
\end{equation}

\vspace{1cm}

\noindent
{\bf \large Acknowledgments}

\medskip

Y.O. would like to thank Volker Schomerus
for useful discussions. We thank Martin Schnabl for informing us of his
independent work on the subject of this paper.
The work of M.K. and B.Z. is supported in part
by the U.S. DOE grant DE-FC02-94ER40818.
The work of M.K. is supported in part
by an MIT Presidential Fellowship.
The work of L.R. is  supported in part
by the National Science Foundation Grant
No.~PHY-0354776.
Any opinions, findings, and conclusions or recommendations
expressed in this material are
those of the authors and do not necessarily reflect the views of
the National Science Foundation.
\medskip

% \appendix

\end{document}